\documentclass[12pt]{iopart}
\usepackage[]{graphicx} 
\expandafter\let\csname equation*\endcsname\relax 
\expandafter\let\csname endequation*\endcsname\relax
\usepackage{amsmath}
\usepackage{amsfonts}
\usepackage{silence} 
\usepackage{subcaption}
\usepackage{hyperref}
\usepackage{color}

\begin{document}

\title[Convergence of GARM]{Improving Convergence of Generalised Rosenbluth Sampling for Branched Polymer Models by Uniform Sampling}
\author{T Roberts$^1$, T Prellberg$^1$}
\address{$^1$ School of Mathematical Sciences, Queen Mary University of London, Mile End Road, London E1 4NS, UK.}
\ead{t.roberts@qmul.ac.uk,t.prellberg@qmul.ac.uk}
\date{\today}

\begin{abstract}
    Sampling with the Generalised Atmospheric Rosenbluth Method (GARM) is a technique for estimating the distributions of lattice polymer models that has had some success in the study of linear polymers and lattice polygons. In this paper we will explain how and why such sampling appears not to be effective for many models of branched polymers. 
    Analysing the algorithm on a simple binary tree, we argue that the fundamental issue is an inherent bias towards extreme configurations that is costly to correct with reweighting techniques.
    We provide a solution to this by applying uniform sampling methods to the atmospheres that are central to GARM. We caution that the ensuing computational complexity often outweighs the improvements gained.
\end{abstract}

\submitto{\JPA}


\section{Introduction}
Over the past twenty years, the Rosenbluth Method and its extensions have been an important tool for the simulation of polymers in equilibrium~\cite{buks_monte_carlo}.
This method, first introduced by Rosenbluth and Rosenbluth in 1955~\cite{rr_sampling}, is a stochastic growth method and as such is an alternative to sampling methods based on Markov chains~\cite{berg_markov_book, landau_binder_book}. One of the advantages of growth methods over Markov Chain methods is that they allow a direct estimate of the number of configurations.

The original Rosenbluth Method suffers from several shortcomings that limit its effectiveness including exponential growth of the variance, and high rates of attrition (samples terminated before reaching the target length)~\cite{batouilis_RR_shortcomings}. Adding pruning and enrichment strategies to this method enables one to overcome these shortcomings, leading to the Pruned-Enriched Rosenbluth Method (PERM)~\cite{grassberger_perm}. PERM can be viewed as a sequential Monte Carlo algorithm with re-sampling. It has been applied to various problems in polymer physics, including homo-polymers, star polymers with fixed topologies, and lattice animals as models for randomly branched polymers~\cite{hsu_grassberger_review}. PERM's ability to produce samples according to any given prescribed weight distribution has made it a method of choice for many problems, although it has also been noted for its potential failures and biases~\cite{hsu_grassberger_review}.

A relevant extension of PERM is the flat histogram stochastic growth algorithm, which adds micro-canonical re-weighting techniques to PERM, enhancing the method's capabilities in polymer simulations, and enabling the probing of more complex energy landscapes~\cite{flatPERM}.

While extensions of the Rosenbluth method have been highly successful for the simulation of linear polymers, there is only limited success in their application to the simulation of branched polymers. It is fairly straightforward to grow these objects using stochastic growth, but there is an added difficulty that configurations are no longer grown in a unique way~\cite{rechnitzer_garm}, a problem which we will refer to as \emph{path degeneracy}. More refined techniques are required to obtain useful results for these models.

A technique to overcome the path degeneracy for tree-like branched polymers is to apply a unique labelling to each configuration and only allow growing trees in ways that maintain the labelling. This ensures that each configuration can only be grown in a single way~\cite{buks_labelled_trees}. The downside of this technique is that finding the valid growth options for a given tree can be very costly in computation time. 

Much work has also been done to sample a cyclic branched polymer model called a \emph{site animal} (precisely defined in section~\ref{sec:models}), by applying strategic re-weighting derived through a connection with percolation clusters~\cite{hsu_grassberger_animals_percolation}. This approach has opened new avenues for exploring complex polymer structures, but it is not readily extensible to other models and has not found extensive use in the literature.

The Generalised Atmospheric Rosenbluth Method (GARM)~\cite{rechnitzer_garm} deals with the issue of over-counting differently. With GARM, one allows for path degeneracy and handles over-counting by adjusting the weights using the concept of \emph{atmospheres}. It is rigorously known that GARM will converge to the correct ensemble average. GARM has been applied to various polymer systems, including self-avoiding walks and self-avoiding polygons~\cite{rechnitzer_garm}. The generality of the GARM technique promises to allow sampling of a broad class of models. 

This paper is structured as follows. Section~\ref{sec:models} contains a careful description of the lattice models, the central notion of atmospheres, the GARM algorithm as applied to these models, and pruning and enrichment techniques.
In section~\ref{sec:garm_results} we use the GARM algorithm to study four common models of branched polymers on the square lattice. We find that the standard implementation of GARM displays a systemic under-estimation of the growth constant for these models. This effect becomes more pronounced as the size of the sampled objects increases.
Section~\ref{sec:binary_trees} analyses GARM applied to a highly simplified model of abstract binary trees. We show that the distribution of sample probabilities has very little overlap with the true distribution of the objects under study. Due to this, the algorithm will often not converge within a useful time frame. As the size of the objects increases the distributions drift further apart. We are able to show that with careful tweaking of the sampling probabilities the algorithm can provide useful results for the simplified model, but this requires knowledge of the target distribution which is not possible for the real lattice animal models.
In section~\ref{sec:flattening} we introduce an approach inspired by the FlatPERM algorithm~\cite{flatPERM}, which we call \emph{atmospheric flattening}; this partitions the state space by the atmospheres of the configurations. It produces reliable results at the cost of increased algorithmic complexity. The paper ends with conclusions in section \ref{sec:conclusions}.

\section{Models and Methods}\label{sec:models}
\subsection{Models of Branched Polymers}
\begin{figure}[ht]
    \centering
    \begin{subfigure}{0.24\textwidth}
        \includegraphics[width=\textwidth]{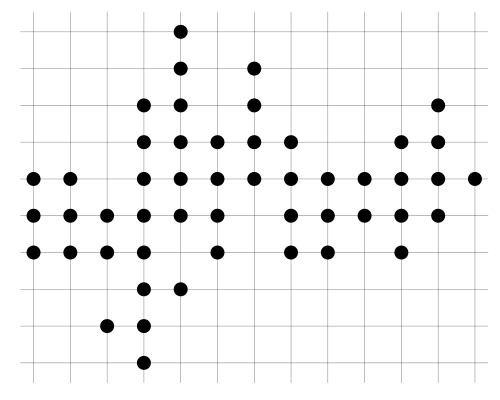}
        \caption{Site Animal}
    \end{subfigure}
    \hfill
    \begin{subfigure}{0.24\textwidth}
        \includegraphics[width=\textwidth]{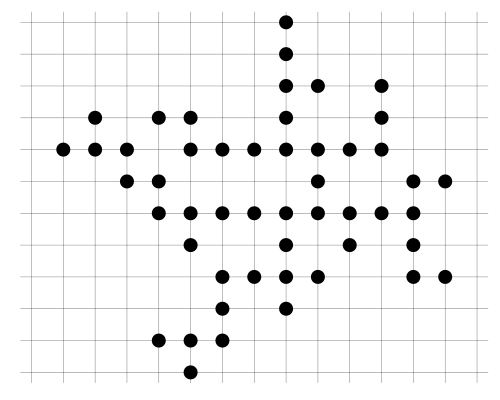}
        \caption{Site Tree}
    \end{subfigure}
    \hfill
    \begin{subfigure}{0.24\textwidth}
        \includegraphics[width=\textwidth]{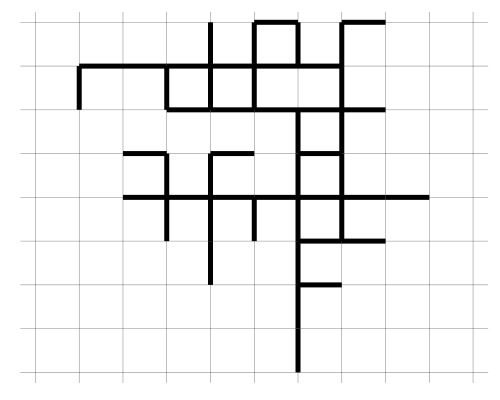}
        \caption{Bond Animal}
    \end{subfigure}
    \hfill
    \begin{subfigure}{0.24\textwidth}
        \includegraphics[width=\textwidth]{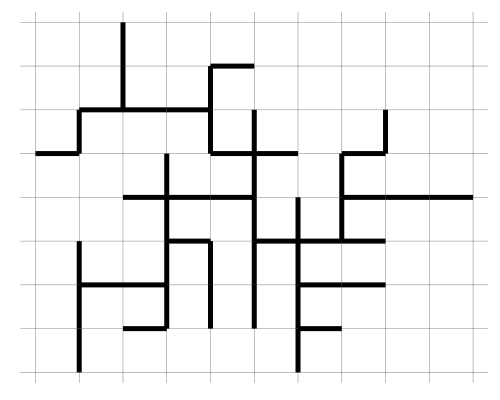}
        \caption{Bond Tree}
    \end{subfigure}
    \caption{Examples of the four models we are studying with size $N=50$. Generated with our sampling code and the Julia Plots package}
    \label{fig:model_pictures}
\end{figure}

We are interested in the properties of four canonical models of branched polymers on the square lattice: site animals, bond animals, site trees, and bond trees. Site animals (also known as Polyominoes in some literature, see for example~\cite{golomb_polyominoes_book, golomb_polyominoes_paper, redelmeier_polyominoes}), are a connected collection of vertices on the lattice, and bond animals are a connected collection of lattice edges. Site and bond trees are the respective subsets of animals which contain no cycles, that is animals with a tree-like structure. We consider these objects up to translations, that is two animals that can be transformed into one-another by a simple translation are considered equivalent. Pictorial representations of these models for the are shown in Figure \ref{fig:model_pictures}. 

Self-interacting versions of these models have been studied previously in~\cite{hsu_grassberger_animals_percolation, derrida_interacting_animals, dickman_interacting_animals}, but throughout this report we will be working solely with the \emph{non-interacting} versions. 

As with the self-avoiding walk~\cite{madras_slade}, it is widely believed that the counting sequences for these objects with respect to $n$, the number of elements, denoted generically as $c_n$, grow asymptotically as
\begin{equation}
\label{eq:counting_seq}
    c_n \sim \mu^n n^{\gamma-1}
\end{equation}
where $\mu$ is the (model-specific) growth constant defined as the limit
\begin{equation}    
    \mu = \lim_{n\to \infty} c_n^{\frac{1}{n}}
\end{equation}

and $\gamma$ is the entropic exponent~\cite{madras_slade}, which is generally believed to be 0 for all four models~\cite{rensburg_book}. 

\subsection{Atmospheres}
Before stating the algorithms, we must first explain the concept of atmospheres. For a configuration $\varphi$, we define the positive atmosphere, denoted $a_+(\varphi)$, as the set of possible successor states that could be grown from $\varphi$ according to the rules of the model. We can similarly define the negative atmosphere, $a_-(\varphi)$, as the set of possible predecessor states from which $\varphi$ could have grown (e.g. for rooted linear polymers the positive atmosphere consists of all unoccupied sites adjacent to the growing end, and the negative atmosphere is the most recently added site). In a helpful abuse of notation, we will also sometimes use $a_+$ and $a_-$ to refer to the size of the respective sets. It will be made clear in context when this is happening. 

For the four models described above, we can define the atmospheres quite straightforwardly as follows. For site trees and bond trees, the positive atmosphere is the set of all sites (bonds respectively), which are adjacent to the tree, and whose addition would not create a cycle. 
The two animal models have a similar positive atmosphere definition, but with the restriction on cycle formation relaxed. 

The negative atmosphere of trees can be defined very simply as the set of leaves of the tree. 
Considering animals to be connected sub-graphs on the lattice, the negative atmosphere of an animal is comprised of all elements in the animal (bonds or sites) whose removal would \emph{not} cause the graph to become disconnected or, equivalently, all elements in the animal which are \emph{not} articulation points (or bridges for bonds) of that sub-graph. 

\subsection{Rosenbluth and GARM}

The Rosenbluth algorithm grows a self-avoiding walk beginning with a single occupied site, usually set to the origin, to which a weight of $1$ is assigned. At each growth step, a random site is chosen from the positive atmosphere and the weight is set to the product of the previous weight and the size of the positive atmosphere at that step. We denote the weight contribution at the step from size $i$ to $i+1$ as $w(\varphi_i \to \varphi_{i+1})$. This process continues until either the desired length , $N$, is reached and the weight is recorded or there are no valid steps to take (i.e. $a^+(\varphi) = \emptyset)$. In the latter case, the walk is discarded without recording, or equivalently given weight $0$. A completed walk $\varphi$ of $N$ bonds thus has a weight equal to the product of the weight contributions:
\begin{equation}
\label{eq:rr_weights}
W(\varphi) = \prod_{i=0}^{N-1} w(\varphi_i \to \varphi_{i+1}) = \prod_{i=0}^{N-1}a_+(\varphi_i)\;.
\end{equation}
By the central limit theorem, the mean weight obtained from $M$ samples converges to the ensemble average,
\begin{equation}
    \frac1M\sum_{m=1}^M \mathbb{P}(\varphi) W(\varphi) \to \sum_{\varphi \in {\varphi}} \mathbb{P}(\varphi)W(\varphi) =\sum_{\varphi \in {\varphi}}1=c_n\;,
\end{equation}
as the weight of a configuration is simply the inverse of the probability. So, by taking the average of the Rosenbluth weights we obtain an unbiased estimator for $c_n$.

\begin{figure}[ht]
    \centering
    \begin{subfigure}{0.47\textwidth}
        \includegraphics[width=\textwidth]{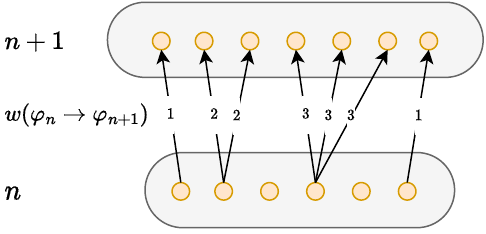}
        \caption{Rosenbluth Sampling. Select uniformly from positive atmosphere. Weight contribution of each transition is $a_+(\varphi_n)$.}
        \label{subfig:rr_diagram}
    \end{subfigure}
    \hfill
    \begin{subfigure}{0.47\textwidth}
        \includegraphics[width=\textwidth]{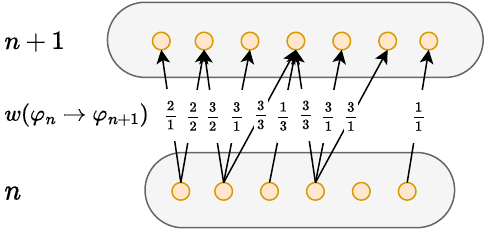}
        \caption{GARM Sampling. Select uniformly from positive atmosphere. Weight contribution of each transition is $a_+(\varphi_n)/a_-(\varphi_{n+1})$.}
        \label{subfig:garm_diagram}
    \end{subfigure} 
    \caption{Diagrammatic representation of Rosenbluth vs GARM sampling, demonstrating how the weights are adjusted to account for path degeneracy. In Rosenbluth sampling the weight applied is the size of the positive atmosphere of the prior state, $a_+(\varphi_n)$, while in GARM it is now the ratio of the prior positive atmosphere size and the successor negative atmosphere size, $a_+(\varphi_n)/a_-(\varphi_{n+1})$. In the case where all states have a single element as the negative atmosphere, $a_-(\varphi_n) = 1 \forall \varphi, n$, the GARM weights reduce to the Rosenbluth weights. This figure is adapted from \cite{rechnitzer_garm}. }
    \label{fig:rr_garm_diagram}
\end{figure}

Figure~\ref{subfig:rr_diagram} shows a diagrammatic representation of the Rosenbluth algorithm. Each small circle represents a specific state (walk) grouped by size. The arrows depict the possible growth steps from one state to another, and the labels on each arrow show the weight contribution for that particular transition.

The weight calculation above only holds when there is a single, unique path through the space of states from the origin to each final state. That is not the case in the naïve approach to sampling branched polymers, as there is path degeneracy in the space of states, but this can be overcome with the ingenious development of GARM. 

GARM proceeds almost exactly as normal Rosenbluth sampling, but with a change to the weight calculation at each step. When taking a step from $\varphi_n$ to $\varphi_{n+1}$ we now define the weight contribution of that step as the ratio of the positive atmosphere at the previous step and the negative atmosphere at the following step, i.e.,
\begin{equation}
w(\varphi_n \to \varphi_{n+1}) = \frac{a_+(\varphi_n)}{a_-(\varphi_{n+1})}\;.
\end{equation}

As there are now multiple paths with potentially different weights that can end in the same state, we can no longer defined a unique weight for a configuration. Instead we define the weight of some particular path from the starting state $\varphi_0$ to a final state $\varphi_N$, which we will denote $\{\varphi_0 \dots \varphi_N\}$. This weight is defined echoing plain Rosenbluth:
\begin{equation}
W(\{\varphi_0 \dots \varphi_N\}) = \prod_{i=0}^{N-1} w(\varphi_i \to \varphi_{i+1}) = \prod_{i=0}^{N-1}\frac{a_+(\varphi_i)}{a_-(\varphi_{i+1})}\;.
\end{equation}
Notice that for self-avoiding walks $a_-(\phi) = 1$ for all configurations and so the weight contribution reduces to the original Rosenbluth weights as stated in Eqn.~(\ref{eq:rr_weights}).
In figure~\ref{subfig:garm_diagram} we present a diagrammatic representation of GARM sampling. As in figure~\ref{subfig:rr_diagram}, the small circles represent states and the arrows represent the possible growth steps. Notice that now there can be multiple arrows pointing into a single state. The labels on the arrows are the weight contribution from the particular transition, which in this case is the GARM weight $a_+(\varphi_n) / a_-(\varphi_{n+1})$.

Let $\{\tau\}$ denote the set of all paths through state space that ends in the state $\varphi$ of size $N$, and let $\mathbb{P}(\tau)$ be the probability of generating this path. The definitions above then ensure that we have 
\begin{equation}
    \sum_{\tau \in \{\tau\}} \mathbb{P}(\tau) W(\tau)  = 1\;,
\end{equation}
and thus the weights generated by GARM are an estimator for $c_n$. A careful proof of this fact can be found in~\cite{rechnitzer_garm}. 

A key concern in the implementation of GARM, as highlighted in~\cite{rechnitzer_garm}, is the efficient computation of atmospheres for a chosen model. As both the positive and negative atmospheres must be used at each step taken in the algorithm, any inefficiencies will be heavily compounded for large runs. 

Fortunately, the positive atmospheres of the four models can be stored in memory and updated on each step of the algorithm in $O(1)$ time complexity. This is also the case for the negative atmospheres of the tree-like models. Unfortunately, the negative atmosphere of the cyclic models is not so trivial. Identifying the articulation points of a graph is a well studied problem in algorithmic graph theory, and is known to have average case complexity $O(V+E)$ for an arbitrary graph with $V$ nodes and $E$ edges~\cite{tarjan_articulations}. In our case both $V$ and $E$ are limited to be $O(n)$, so we can expect the calculation of the negative atmosphere for bond and site animals to take $O(n)$ time at each step of the algorithm.


\subsection{Pruning and Enrichment}

A key improvement to the Rosenbluth algorithm is the Pruned and Enriched Rosenbluth Method (PERM). PERM adds a new step to the algorithm after growing and calculating the weight, but before the recording of the weight. In this step samples with low weight are terminated early, called \emph{pruning}, and walks with high weight are split into several copies of more reasonable weight, called \emph{enrichment}. By pruning and enriching in a well chosen way, PERM provides a noticeable reduction in variance, and for the self-avoiding walk model mitigates the effects of attrition that prevent effective sampling of large walks.

We choose to use the technique of continuous pruning and enrichment proposed in \cite{flatPERM}. Let $r$ be the ratio of the current configuration's weight, $W(\varphi_n)$, and the running average of all weights sampled at size $n$, $\hat{W}_n$: 
\begin{equation}\label{eq:enrichment_ratio}
    r = \frac{W(\varphi_n)}{\hat{W}_n}
\end{equation}
We want to prune/enrich such that the expected number of copies is $r$. We accomplish this by the following method:

If $r \leq 1$ then prune. Draw a random number $p \sim U(0,1)$. If $p\leq r$ then terminate the sample, recording the weight as 0. If $p>r$ keep the sample and record the weight of the sample as $\hat{W}_n$. 

If $r > 1$, then enrich. Draw a random number $p \sim U(0,1)$. If $p \leq r\bmod 1$ make $\lfloor r \rfloor$ copies of $\varphi_n$, or if $p > r\bmod 1$ make $\lceil r \rceil$ copies. Record the weight of each copy as $\hat{W}_n$. Each copy is continued independently according to the rules of the algorithm.

Note that while it is pedagogically helpful to consider pruning and enrichment separately, algorithmically the case $r\leq1$ can be subsumed in the case $r>1$ as enrichment with zero or one copy.

When discussing pruned and enriched sampling we use the word \emph{tour} to mean the collection of all samples grown from a single start, including all enrichment copies at all steps. We also define the \emph{tour weight} to be the sum of all weights sampled at the target length $N$ within a single tour. An important consideration of this method is that samples within a single tour can potentially be highly correlated reducing the effective sample count.

\section{GARM and pruned-enriched GARM}\label{sec:garm_results}

\subsection{Simulation Results}

We sampled the four models of branched polymers described above using GARM with and without pruning and enrichment up to size $N=200$. We gathered data from $1,000,000$ GARM samples and $100,000$ Pruned-Enriched GARM tours for each model. We used a parallel implementation of GARM inspired by~\cite{parallel_perm}. This technique involves running multiple tours in parallel with mutually interacting weight estimates. It has been found to increase the convergence rate of PERM compared to naive parallelisation, and allows for much more effective use of the large number of computing cores available in modern HPC environments.

The approximate counting sequences produced by these simulations can be used to estimate the growth constant of each model. We follow the procedure in~\cite{madras_slade} as follows. By taking logarithms in Eqn.~(\ref{eq:counting_seq}) we get a linear expression in two variables, on which we can perform a regression analysis to obtain estimates of the growth constant, $\mu$, and the entropic exponent, $\gamma$. In what follows, it suffices to focus on $\mu$. More sophisticated series analysis techniques exist, but for the purposes of this paper, the simple regression proves to be sufficient. 

\begin{figure}[ht]
    \centering
    \begin{subfigure}{0.42\linewidth}
        \includegraphics[width=\linewidth]{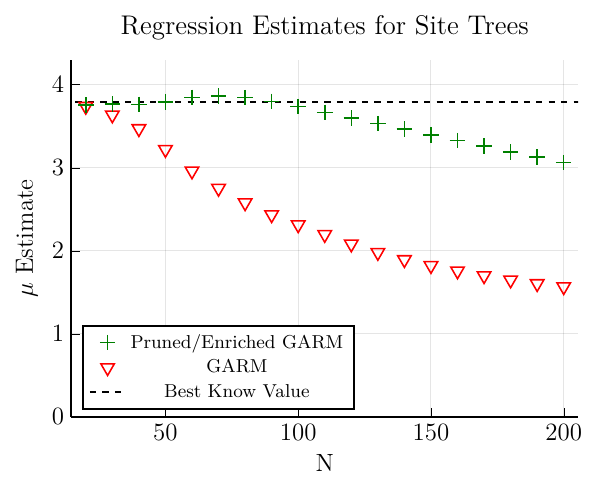}
    \end{subfigure}
    \begin{subfigure}{0.42\linewidth}
        \includegraphics[width=\linewidth]{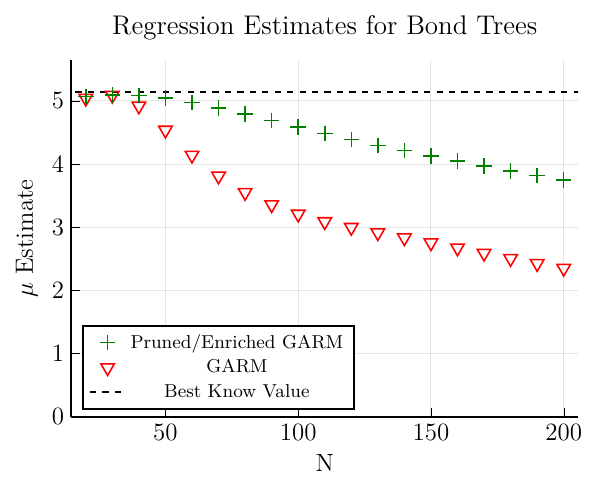}
    \end{subfigure}
    
    \begin{subfigure}{0.42\linewidth}
        \includegraphics[width=\linewidth]{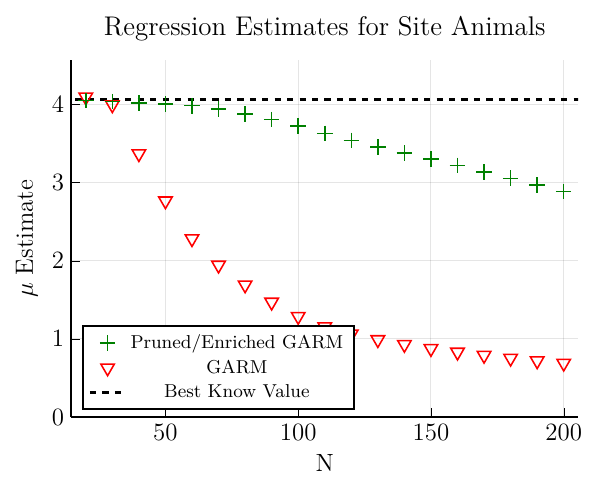}
    \end{subfigure}
    \begin{subfigure}{0.42\linewidth}
        \includegraphics[width=\linewidth]{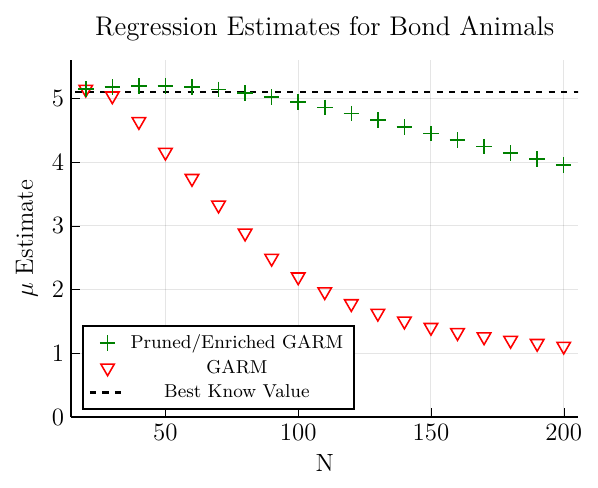}
    \end{subfigure}
    \caption{Estimates of the growth constants obtained by regression from the weight sequence from GARM with and without pruning and enrichment, together with the best known values, up to $N=200$.}
    \label{fig:mu_estimates}
\end{figure}

Figure~\ref{fig:mu_estimates} summarises our findings for the four models. It plots estimates of the growth constant obtained by these regression techniques over a window $[0,N]$. We show the results for both GARM and Pruned-Enriched GARM on each plot, along with a guide for the eye at the current best known value of the growth constant for each model \cite{jensen_animals_and_trees,gaunt_bond_trees,mertens_bond_animals}.

It is clear from all four panels that GARM systematically underestimates the growth constant as the system size increases. This is contrary to what one might expect of an increasing regression window, where more terms in the sequence should result in more accurate parameter estimates. Pruning and enrichment alleviates this effect to an extent, seemingly providing reliable estimates up to $N\approx50$, but in all cases the estimates systematically tail off once $N$ passes $100$.

The fact that the Rosenbluth-like method struggles will not be surprising to those familiar with its application to linear polymers. One of the  to using the Rosenbluth method to study linear polymers is the exponential attrition of self-avoiding walks. That is, it becomes exponentially likely to hit a dead-end as the walk grows, terminating the sampler before it reaches target length~\cite{madras_slade, hsu_grassberger_review}. 
The other well-known problem with plain Rosenbluth, which it shares with other sequential importance sampling techniques~\cite{delmoral_sequential_MC}, is the exponential growth of the sample variance. As larger objects are probed, it becomes increasingly likely that the weight of a very small number of samples will dominate the average, being several orders of magnitude higher than any others. In this case it can take an impractically long time to generate sufficiently many samples that the sample mean will converge to the expectation value~\cite{batouilis_RR_shortcomings, hsu_grassberger_review}, despite the proven fact that it will eventually converge.

For linear polymers, the addition of pruning and enrichment solves both of these problems. Sampling SAWs with PERM results in roughly constant numbers of samples at each size (although, as noted, due to correlation effects the effective sample count may be lower). The sample variance is also well controlled: low-weight samples are pruned, and high-weight configurations are enriched, with the weights being re-balanced at each step~\cite{grassberger_perm}.

For the branched polymer models we study, the positive atmosphere will always be strictly greater than zero for any configuration. Therefore there is no attrition, and the first problem is not relevant for these models. It is therefore reasonable to expect that the sample variance leads to systematic underestimation for GARM.
Pruning and enrichment reduces this sample variance, improving the growth constant estimates significantly for small system sizes. However, there is still a definite tail-off for for large $N$ contrary to expectations. This suggests that there is some deeper problem at work than just those mentioned above.

\subsection{Tourweight Diagnostics}

In~\cite{hsu_grassberger_review}, Hsu and Grassberger note that depending on the amount of pruning and enrichment taking place, the degree of correlation within a tour `could be so strong as to render the method obsolete.' The sample average weight ends up dominated not by a single sample, but by the (highly correlated) samples from a single tour, effectively reverting to the problem of Rosenbluth sampling, but with tours replacing samples. 

\begin{figure}[ht]
    \centering
    \begin{subfigure}{0.45\textwidth}
        \includegraphics[width=\linewidth]{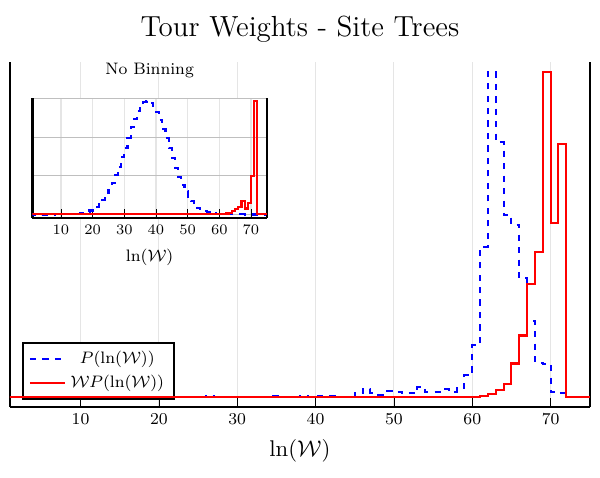}
    \end{subfigure}
    \begin{subfigure}{0.45\textwidth}
        \includegraphics[width=\linewidth]{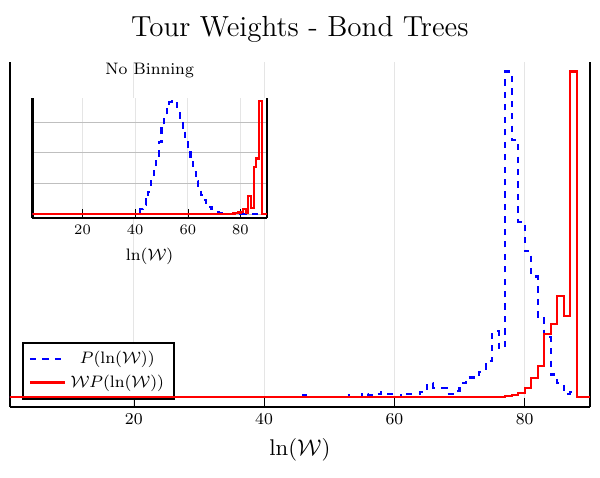}
    \end{subfigure}
    \begin{subfigure}{0.45\textwidth}
        \includegraphics[width=\linewidth]{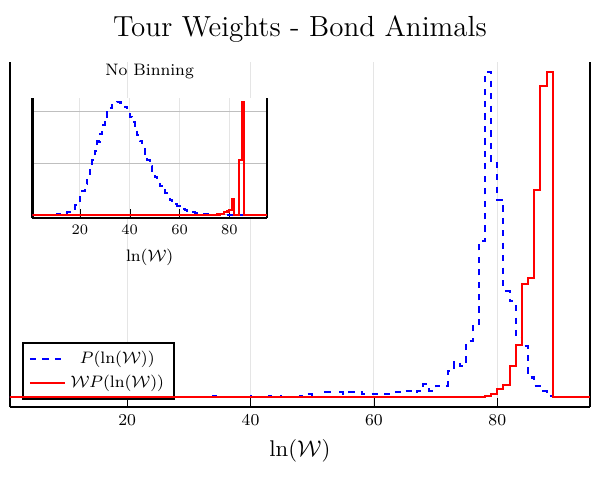}
    \end{subfigure}
    \begin{subfigure}{0.45\textwidth}
        \includegraphics[width=\linewidth]{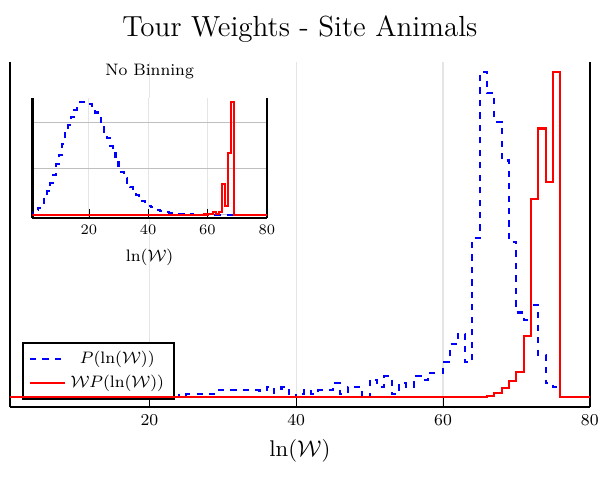}
    \end{subfigure}
    
    \caption{Diagnostic tools inspired by~\cite{hsu_grassberger_review}. The dashed blue line is a histogram of $P(\ln(\mathcal{W}))$, and shows where in the weight distribution the bulk of samples are taken. The solid red line is the weighted distribution, $\mathcal{W}P(\ln(\mathcal{W}))$, and should serve as an estimator for the true distribution. Both distributions are scaled such that their peaks are at the same height. Data are presented for sizes $N=50$.}
    \label{fig:toursize_diagnostics}
\end{figure}

Grassberger suggests the use of \emph{tour weights} as a diagnostic~\cite{grassberger_tourweight_comment}. By plotting the distribution of the logarithms of tour weights, $P(\ln(\mathcal{W}))$, alongside the weighted distribution $\mathcal{W}P(\ln(\mathcal{W}))$ and considering the overlap, we get an effective heuristic with which to study this issue. $P(\ln(\mathcal{W}))$ shows where in the weight distribution the bulk of samples are taken. $\mathcal{W}P(\ln(\mathcal{W}))$ should serve as an estimator for the true distribution. However, in the case where there is poor overlap, the weighted distribution is used more as a proxy for the relative location of the true distribution, much of which may not have been sampled at all. Good overlap between the weighted and unweighted distributions suggests that the samples \emph{may} produce reliable data (although this is not a guarantee). If the distributions have little overlap, however, this signals that the results have been dominated by a small number of tours, and will likely be unreliable. 

In figure~\ref{fig:toursize_diagnostics} we present this diagnostic for each of the four models with data from runs up to $N=50$. In each case, the dashed blue line is the unweighted distribution, $P(\ln(\mathcal{W}))$, and the solid red line is the weighted distribution, $\mathcal{W}P(\ln(\mathcal{W}))$.
The four panels show the distributions with pruning and enrichment applied, and the respective insets are for runs of GARM without pruning and enrichment.

In the plots without pruning and enrichment, the ``tourweight'' distribution is equivalent to the distribution of sampled weights, $P(\ln(W))$, since each ``tour'' has only one sample. In these plots $WP(\ln(W))$ clearly sits far to the right of $W$ with no significant overlap, neatly demonstrating the problem of exponential variance in Rosenbluth sampling. When adding pruning and enrichment, the sampling distribution has clearly been skewed towards the weighted distribution: there is now a non-negligible amount of overlap between the two, but in all four cases the peak of $P(\ln(\mathcal{W}))$ falls in a region where $\mathcal{W}P(\ln(\mathcal{W}))$ is small. 
From these plots we can gather that poor distributions of tour weights very likely play a contributing factor in the underestimation of growth constants with GARM, even with the use of pruning and enrichment.

\subsection{Atmospheric Parameter Space}

To gain further insight into the sampling process, we constructed contours that visualize the sample count and average weight relative to the positive and negative atmospheres. These contours provide a comprehensive representation of the sampling distribution. As with the previous diagnostic method, this visualisation allows us to probe where the majority of samples are taken relative to the bulk of the ensemble's weight, but with the added benefit that we can now see this information relative to the key parameters of the algorithm: the atmospheres.

\begin{figure}[ht]
    \centering
    \begin{subfigure}{0.45\textwidth}
        \includegraphics[width=\linewidth]{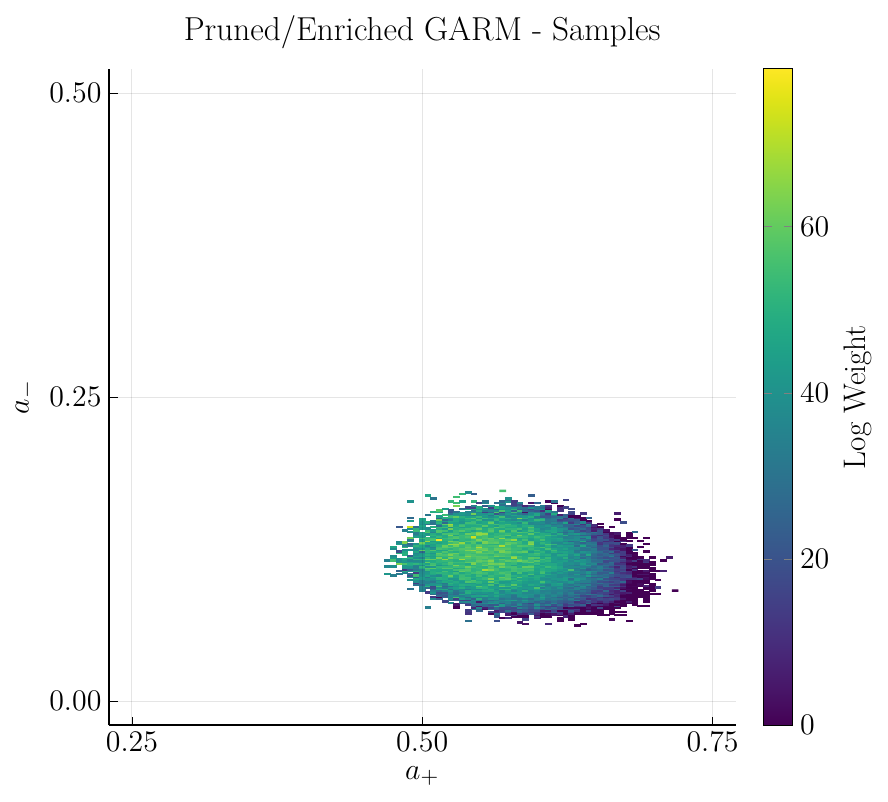}
    \end{subfigure}
    \begin{subfigure}{0.45\textwidth}
        \includegraphics[width=\linewidth]{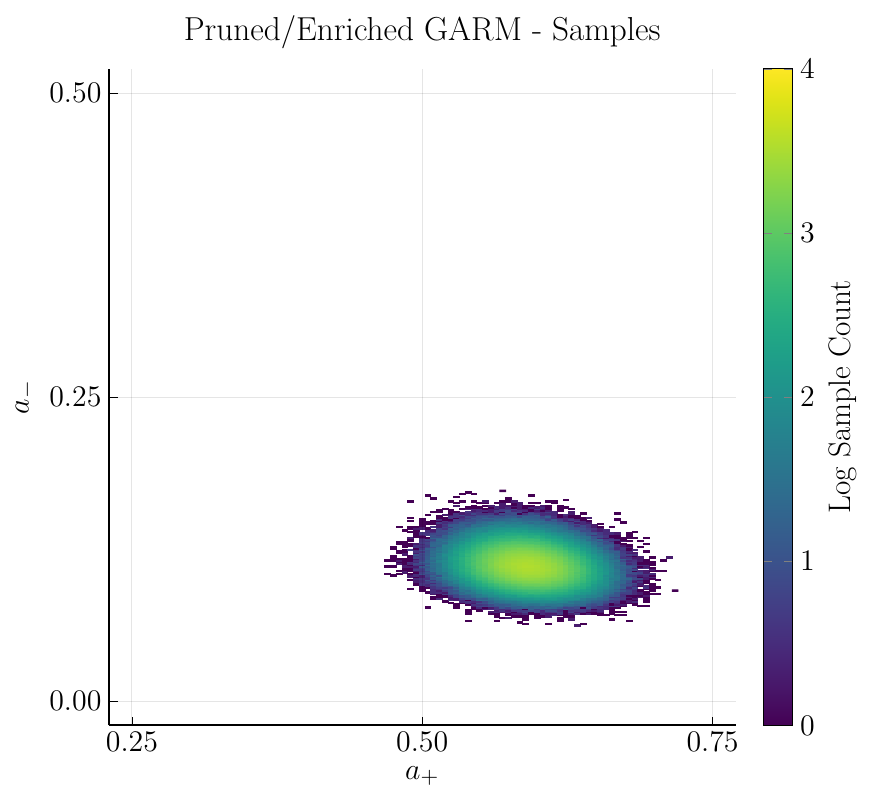}
    \end{subfigure}
    \begin{subfigure}{0.45\textwidth}
        \includegraphics[width=\linewidth]{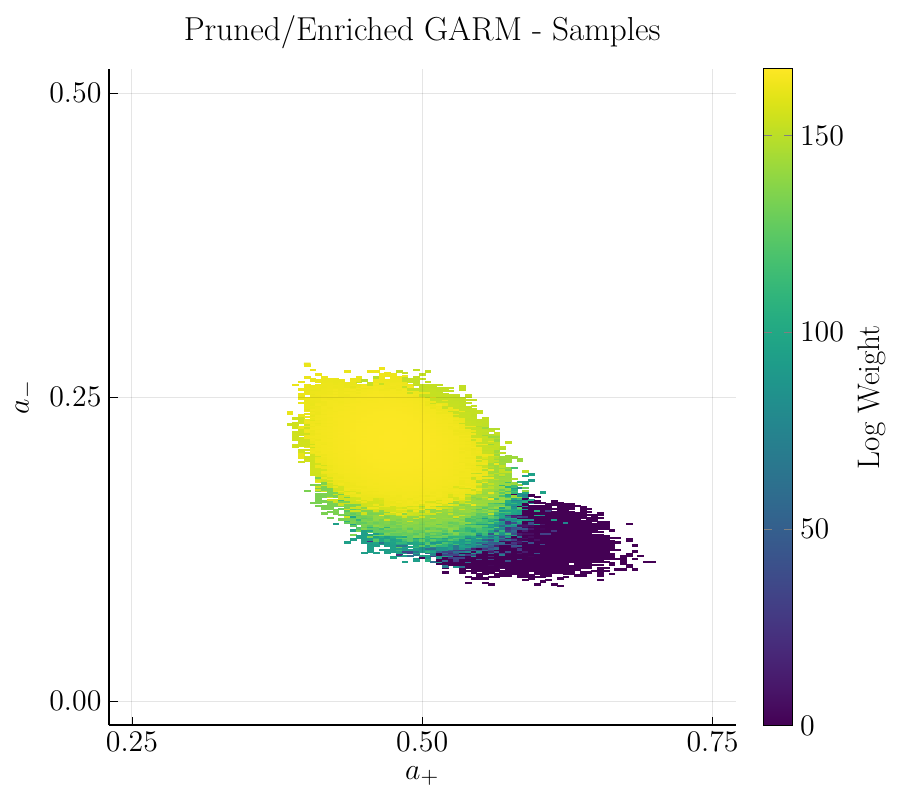}
    \end{subfigure}
    \begin{subfigure}{0.45\textwidth}
        \includegraphics[width=\linewidth]{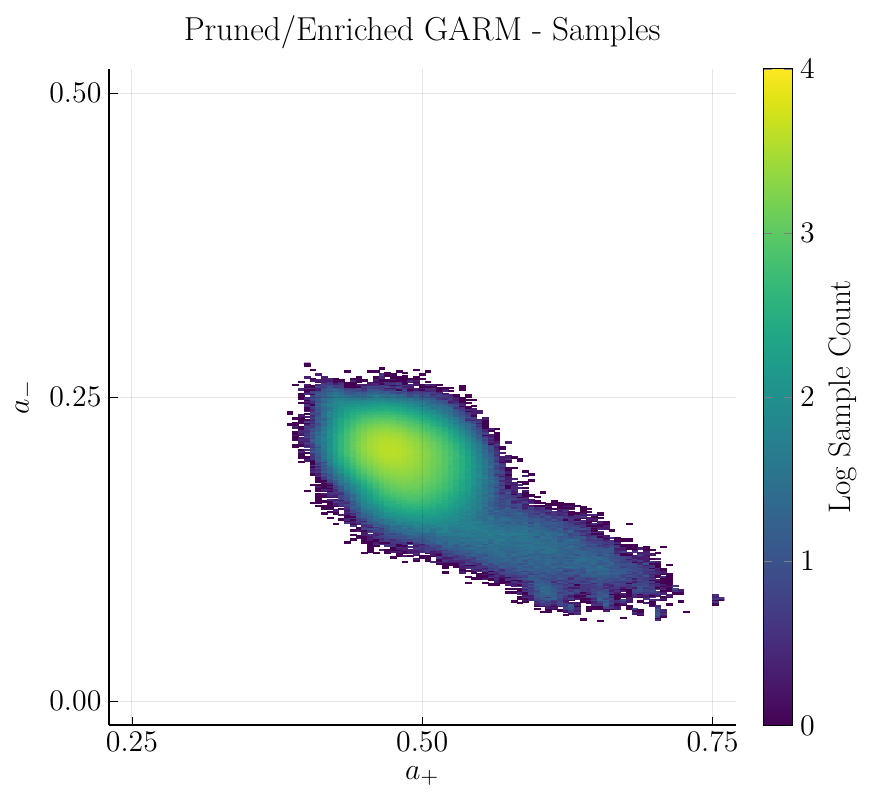}
    \end{subfigure}
    \caption{Atmospheric contours of weights (left) and samples (right) for site trees. The left-hand plots show the weights, and the right hand figures show the sample counts. The top two panels are sampled using plain GARM, and those in the bottom row are using Pruning and Enrichment. The x- and y-axes are the positive and negative atmosphere sizes respectively, and have been normalized such that values lie in the range $[0,1]$. The plot has been cropped to focus on the sampled region. In all cases the data is shown for size $N=400$.}
    \label{fig:pegarm_contours}
\end{figure}

Figure~\ref{fig:pegarm_contours} shows these contours for site trees (the scenario for the other three models is very similar). The top row shows weight distributions (left) and sample distributions (right) for GARM. The peak on the samples plot is in an area of relatively lower weight, and from the apparent weight gradient increasing from right to left across the area sampled, it is likely that the peak of the true weight distribution lies to the left of the sampled area. We can draw the conclusion that while the average sampled weight provably converges to the true counting numbers, in practice we are extremely unlikely to ever sample from some of the more statistically relevant configurations, leading to extremely long convergence times.

The bottom row of contours are for GARM with added pruning and enrichment. We note that the sampling region has shifted significantly with respect to plain GARM, pulling the sampler towards the higher-weight region of configuration space. However, the samples and weights peaks still lack significant overlap, indicating that the true distribution is still not well sampled.

In summary, sampling branched polymers with GARM suffers from issues beyond those seen when sampling linear polymers. While for linear polymers pruning and enrichment is able to control variance growth, for branched polymers pruning and enrichment alone is no longer sufficient. It does not by itself adequately adjust the sampling region towards the true distribution, leading to the issues seen with the tourweight statistics and growth constant estimates.

As a consequence, attempts to extract growth constant estimates results in significant and systemic underestimation, which persists even when larger objects are examined. 
These limitations restrict the ability of the algorithm to obtain meaningful data even up to sizes of $N\approx50$. For the models we are studying, objects of this size can still be enumerated exactly using cutting edge techniques like those found in~\cite{jensen_animals_and_trees}. Considering that the GARM counts would introduce additional Monte Carlo error alongside the systematic error caused by poor quality sampling, any confidence interval we could assign to our estimates would be significantly less competitive compared to the current state of the art.

\section{Binary Trees: A Productive Toy Model}\label{sec:binary_trees}

\subsection{Rooted, Ordered Binary Trees}

In order to better understand the source of the problems encountered when sampling lattice animals with GARM, we turn our attention to a simpler model: rooted, ordered\footnote{By ordered we mean that a left child and a right child are two different configurations}, binary trees. These trees are enumerated by Catalan numbers
\begin{equation}
    C_n = \frac{1}{n+1}\binom{2n}{n}\;.
\end{equation}
One can also enumerate these by both size, $n$, and number of leaves, $k$, with the formula 
\begin{equation}
T_{n,k} = \frac{2^{n+1-2k}}{n} \binom{n}{k} \binom{n-k}{k-1}\;.
\end{equation}
The derivation of $T_{n,k}$ is fairly straightforward from their two-variable generating function, $T(x,y)$, which satisfies $T(x,y) = xy + 2x T(x,y) + x T(x,y)^2$ using Lagrange inversion or other methods.\footnote{This is equivalent to a Dyck path counting problem, see OEIS sequence \href{https://oeis.org/A091894}{A091894}.}

We can define atmospheres for these binary trees in an obvious way. Each node in the tree can have two children, so there is an element of the positive atmosphere for each unoccupied child. For all binary trees of this type with $n$ nodes, $a_+(n)=n+1$ (as can also be easily seen by induction). The negative atmosphere is then just the set of leaves of the tree, so that $a_-(n) \in [1, \lceil n/2 \rceil ]$. Since $a_+$ is fixed by n, we only have one free parameter in our sampling: $a_-$.

\begin{figure}[ht]
    \centering
    \includegraphics[width=0.8\textwidth]{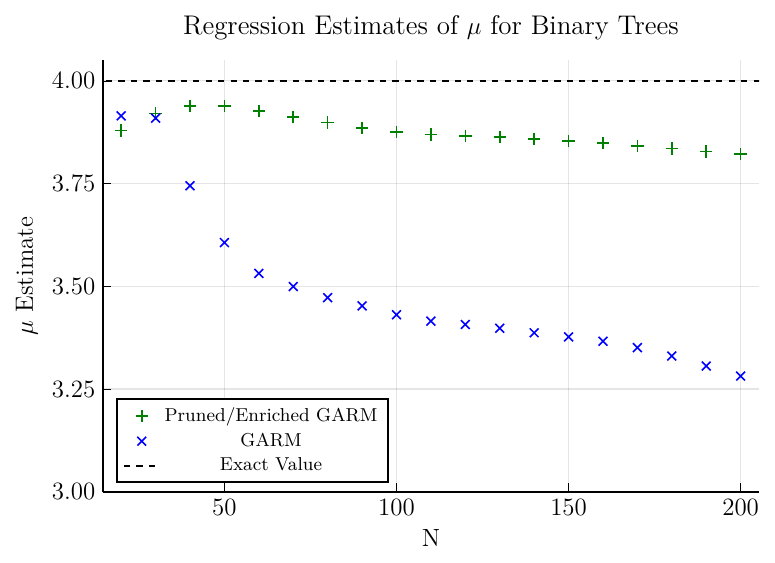}
    \caption{Estimates of the growth constant of ordered, rooted binary trees obtained by regression from the weight sequence from GARM with and without pruning and enrichment, together with the true known value $4$, up to $N=200$, similar to figure~\ref{fig:mu_estimates}.}
    \label{fig:binarytree_results}
\end{figure}

Before we present a theoretical analysis of GARM through the lens of binary trees, it is worthwhile to demonstrate that despite being a drastically simplified model, the estimates obtained when sampling them using GARM exhibit the same sorts of problems seen for the more complex models. Figure~\ref{fig:binarytree_results} shows a version of the regression estimate figure~\ref{fig:mu_estimates} from the previous section replicated for binary trees. For this model it is well known that the asymptotic growth constant is exactly $4$, and this is marked as the dashed line on the plot.

This plot shows the same systematic underestimation as seen with the lattice animal models. Pruning and enrichment show definite improvement, but there is still a clear bias towards lower values, which worsens as $N$ increases. It is quite apparent that binary trees see the same decay in growth constant estimates. We believe that this justifies the assumption that binary trees make a useful model to probe the details of these failures.

\subsection{GARM as a Stochastic Process}

Now, given a binary tree (of the type above) with $n$ nodes and $k$ leaves, choose an element uniformly at random from the positive atmosphere and append it to the tree to create a new tree of size $n+1$. Consider, what are the possible values of $a_-$ for the new tree, and with what probabilities will they occur?

To answer this question, notice that the elements of $a_+$ for any tree can be divided into two distinct categories: children of leaves and children of unsaturated branches. By branch we mean nodes with children and an unsaturated branch is one with only a single child. For a node chosen from the first category adding it to the tree would remove the parent from the negative atmosphere and add itself, for a net change of $0$. Furthermore, each leaf node has two potential children, so there are $2k$ options to grow in this way. 

In the second case (appending to a branch), the parent is not an element of the negative atmosphere, so there will be a net change to $a_-$ of $+1$ with $n+1-2k$ possible choices of this form. We can therefore either keep the same number of leaves or increase by $1$ with transition probabilities
\begin{equation}
\mathbb{P}(k_n \to k_{n+1}) = 
    \begin{cases}
    \frac{2k}{n+1}      & \text{if } k_{n+1}=k_n\;, \\
    \frac{n+1-2k}{n+1}  & \text{if } k_{n+1}=k_n +1\;.
    \end{cases}
\end{equation}

Using these transition probabilities we can derive a recursive formula for the probability of obtaining a tree with $n$ nodes and $k$ leaves, starting from the tree consisting only of a single node with no children and growing using this process. We will denote this function as $P(n,k)$, which is given by the recurrence 
\begin{equation}
P(n,k) = \frac{2k}{n+1} P(n-1,k) + \frac{n+1-2k}{n+1} P(n-1,k-1)
\end{equation}
with boundary conditions
\begin{equation}
P(1,1) = 1 \text{; }
P(n,0) = 0 \text{; }
P(n, \lceil n/2 \rceil + 1) = 0
\end{equation}
for all n.

\begin{figure}[ht]
    \centering
    \begin{subfigure}{0.32\textwidth}
        \includegraphics[width=\textwidth]{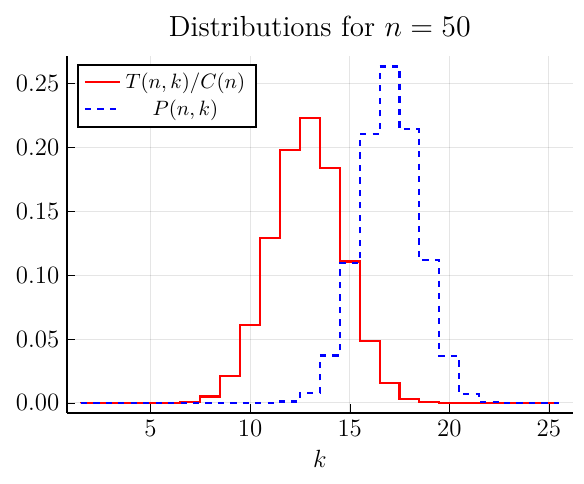}
    \end{subfigure}
    \hfill
    \begin{subfigure}{0.32\textwidth}
        \includegraphics[width=\textwidth]{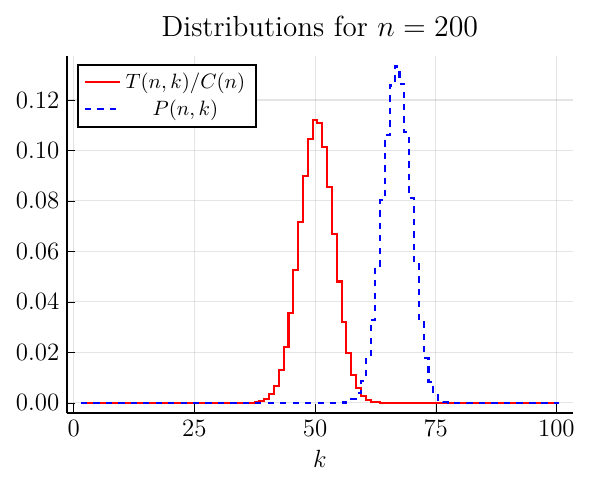}
    \end{subfigure}
    \hfill
    \begin{subfigure}{0.32\textwidth}
        \includegraphics[width=\textwidth]{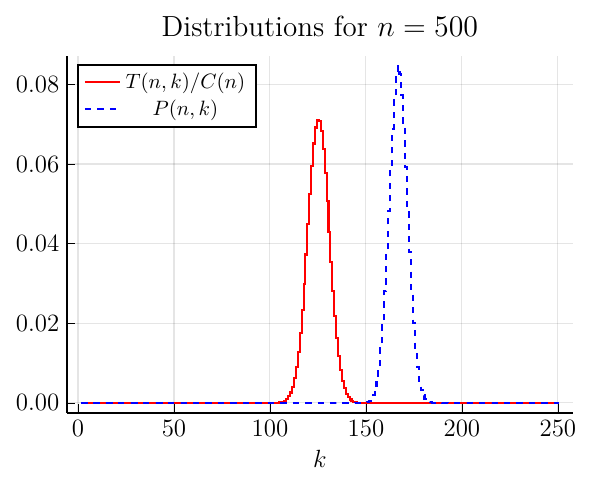}
    \end{subfigure}
    \caption{Distribution of samples taken with GARM (dashed blue) against the actual distribution of binary trees enumerated by size, $N$, and number of leaves, $k$ (solid red). At size $N=50$ there is an acceptable amount of overlap for importance sampling, but  As $N$ increases the distributions narrow and the overlap decreases to almost nothing. This would suggest that using $P(n,k)$ as an importance distribution to sample from $T_{n,k}$ will have poor convergence.}
    \label{fig:bintree_sample_distributions}
\end{figure}

The key insight here is that the function $P(n,k)$ represents the importance distribution when sampling rooted, ordered, binary trees using GARM. Deriving a closed-form expression for this recursive formula is non-trivial\footnote{When normalized as $P(n,k) / (n+1)!$, this function also counts the permutations of n elements with k peaks. See OEIS sequence \href{https://oeis.org/A008303}{A008303} for more detail.}.
Using numerical methods to generate the terms of the sequence and comparing this to what we know to be the ``true" population distribution from above (normalized as $T_{n,k}/ C_{n}$), we were able to observe the trend shown in figure \ref{fig:bintree_sample_distributions}. The sample distribution sits to the right of the true distribution, and while there is some overlap at size $N=50$, we already see that the peak of the true distribution is no longer well sampled by the sample distribution. The distributions narrow further as $N$ increases and the amount of overlap tails off rapidly. For importance sampling to work we would expect these two distributions to maintain sufficient overlap for all $n$, which is clearly not the case.

\begin{figure}[ht]
    \centering
    \includegraphics[width=0.8\textwidth]{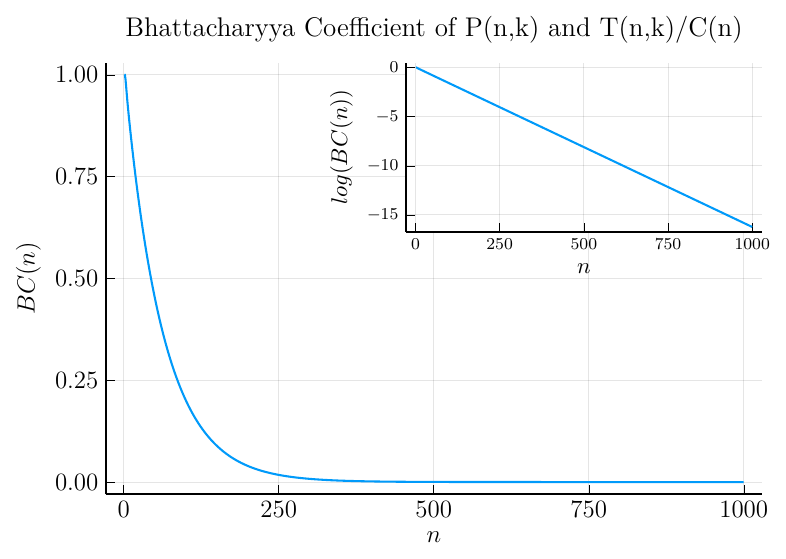}
    \caption{Plot of the Battacharyya Coefficient for the sampling and true distributions of binary trees of size $n$ relative to the number of leaves $k$. The value decays rapidly to zero implying that there is no effective overlap between the distributions for large $n$. The inset has a log scale showing the exponential decay rate.}
    \label{fig:BC_binarytrees}
\end{figure}

We can also look more quantitatively at the difference between distributions with the use of the Bhattacharyya Coefficient: a measure of the amount of overlap between two distributions~\cite{bhattacharyya}. It is defined for two distributions $P$ and $Q$ on the same domain $\mathcal{X}$ as
\begin{equation}
    BC(P, Q) = \sum_{x \in \mathcal{X}} \sqrt{P(x) Q(x)}\;.
\end{equation}
We can calculate the Bhattacharyya Coefficient for the sampling and true distributions of binary trees at some size $n$ as
\begin{equation}
    BC(n) = \sum_{k \in [1, \lceil n/2 \rceil]} \sqrt{P(n,k) \frac{T_{n,k}}{C_n}}
\end{equation}
and have plotted the value for a range of $n$ in figure~\ref{fig:BC_binarytrees}. The plot makes more precise what we had observed in figure~\ref{fig:bintree_sample_distributions}, that the amount of overlap between the distributions becomes significantly worse as $n$ increases. Further numerical analysis of the data in this plot reveals that not only does $BC(n)$ decay to zero, but that it does so exponentially fast in $n$, as shown in the inset plot. 

\subsection{Corrective Measures: Biasing the Sampling Probabilities}
Recognizing the systemic issues with sampling of binary tree configurations using GARM, it would be useful if one could modify the transition probabilities in such a way that the distributions maintain a good degree of overlap. That is, we would like to find a method of biasing the sampling probabilities such that the distribution of samples produced by the algorithm more accurately reflect the true distribution of binary tree. In the 2-parameter enumeration, this would mean generating $(n,k)$ with probability $P(n,k) = T_{n,k} / C_n$. We therefore want to find a bias function $f(n,k)$, satisfying 
\begin{equation}
P(n,k) = f(n-1,k) P(n-1,k) + (1- f(n-1,k-1)) P(n-1,k-1)\;,
\label{eq:probability_recursion}
\end{equation}
which can be solved for $f(n,k)$ to obtain the explicit bias function
\begin{equation}    
f(n,k) = \frac{n+2k}{2n+1}\;.
\end{equation}

With this we can design a new process. We choose from all possible growth steps such that the probability of the value of the negative atmosphere remaining unchanged is $f(n,k)$ and the probability of it increasing by one is $1-f(n,k)$. As noted above, there are $2k$ ways of growing $k\to k$ and $n+1-2k$ ways of growing $k \to k+1$. We can choose uniformly within these categories, and thus the probability of making the transition $\varphi_n \to \varphi_{n+1}$ should be
\begin{equation}
\mathbb{P}(\varphi_n \to \varphi_{n+1}) = 
\begin{cases}
    \frac{n+2k}{2k(2n+1)}   & \text{if } a^-(\varphi_{n+1}) = a^-(\varphi_n) = k\;, \\
    \frac{1}{2n+1}          & \text{if } a^-(\varphi_{n+1}) = a^-(\varphi_n) + 1 = k+1\;.
\end{cases}
\end{equation}

As we have changed the probabilities of growing configurations, we must also adjust the weights to compensate. As with normal GARM, the weight of a sequence of configuration, $W(\{\varphi_0 \dots \varphi_n\})$, is comprised of the product of weight contributions from each transition $w(\varphi_k \to \varphi_{k+1})$. However, since we have changed the probability of sampling a particular configuration, we must also change the weights to compensate. The weight contribution at each step is now
\begin{equation}
w(\varphi_n \to \varphi_{n+1}) = \frac{1}{a^-(\varphi_{n+1}) \mathbb{P}(\varphi_n \to \varphi_{n+1})}\;.
\end{equation}

It is important that we still divide by the negative atmosphere to account for multiple-counting due to path degeneracy. Inserting the probabilities above we get for the weight contributions in the reduced representation parameterised by $(n,k)$:
\begin{equation}
w((n,k_n) \to (n+1, k_{n+1})) = 
\begin{cases}
    \frac{4n+2}{n+2k}       & \text{if } k_{n+1} = k_n = k\;, \\
    \frac{2n+1}{k+1}        & \text{if } k_{n+1} = k_n + 1 = k+1\;.
\end{cases}
\end{equation} 

\begin{figure}[ht]
    \centering
    \includegraphics[height=5.2cm]{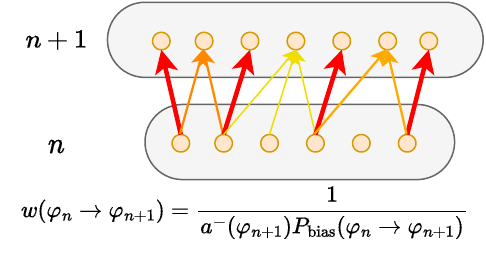}
    \caption{A diagrammatic representation of the biasGARM algorithm, showing how the probability of selecting a given path is now non-uniform (denoted by arrow color). The weight contribution of a transition from $\varphi_n$ to $\varphi_{n+1}$ is now dependent on the biased probabilities. This figure is adapted from figure~\ref{fig:rr_garm_diagram} which is adapted from \cite{rechnitzer_garm}.}
    \label{fig:biasgarm_diagram}
\end{figure}

In figure~\ref{fig:biasgarm_diagram} we show an updated form of the diagrams from figure~\ref{fig:rr_garm_diagram}. We use the color of the arrow to represent the probability of making each transition.

\begin{figure}[ht]
    \centering
    \includegraphics[width=0.8\textwidth]{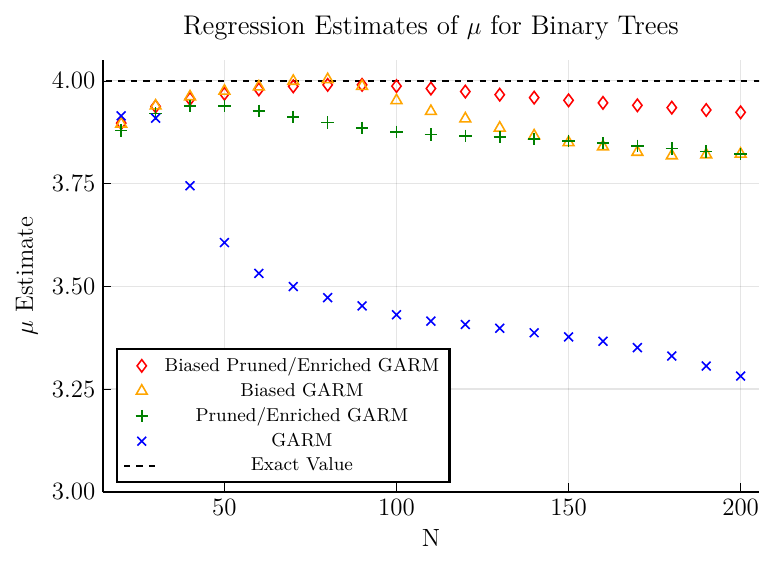}
    \caption{Results obtained from sampling binary trees using "perfect" biasing, with and without pruning and enrichment. The biasing procedure does have a noticeable improvement up to about size $100$, but then we still see a tail-off to around the quality of unbiased samples with pruning and enrichment by size $200$. }
    \label{fig:biasgarm_results}
\end{figure}

We use this new biasing strategy to sample Binary Trees with GARM. As with the lattice animal models, we gathered data from $1,000,000$ GARM samples and $100,000$ Pruned-Enriched GARM tours. The results are shown in Figure \ref{fig:biasgarm_results}. Even without pruning and enrichment we observe estimates closer to the known values. This supports our hypothesis that the bad importance distribution is indeed a significant contributor to the poor quality of results we were obtaining. 

Unfortunately, while this "perfect" bias function works for the binary tree model, it is impractical for more complex models. For one, it requires precise knowledge of the target distribution, the lack of which is the main motivation for studying these models with Monte Carlo methods in the first place. Moreover, the coupling of the positive atmosphere with $n$ in binary trees simplifies the problem significantly. For lattice models, the lattice embedding of the trees destroys this coupling and adds another layer of complexity to the issue. Thus, while the introduction of a bias function provides an interesting corrective measure for the shortcomings of the GARM algorithm, it is not a universally applicable solution. 

\subsection{Flat Sampling by Biasing}
Rather than attempt to mimic the exact distribution, we might instead consider trying to sample uniformly in leaf count. This can be accomplished with a bias function satisfying the specialized form of Eqn. (\ref{eq:probability_recursion}):
\begin{equation}
\frac{1}{\lceil n/2 \rceil} = f(n-1,k) \frac{1}{\lceil (n-1)/2 \rceil} + (1 - f(n-1,k-1)) \frac{1}{\lceil (n-1)/2 \rceil}\;,
\end{equation}
which can be solved by cases to get the following formula:
\begin{equation}
f(n,k) = 
\begin{cases}
    \frac{n+2-2k}{n+2}  & \text{if } n \text{ is even}\;,\\
    1                   & \text{otherwise}\;.
\end{cases}
\end{equation}

\begin{figure}[ht]
    \centering
    \includegraphics[width=0.8\textwidth]{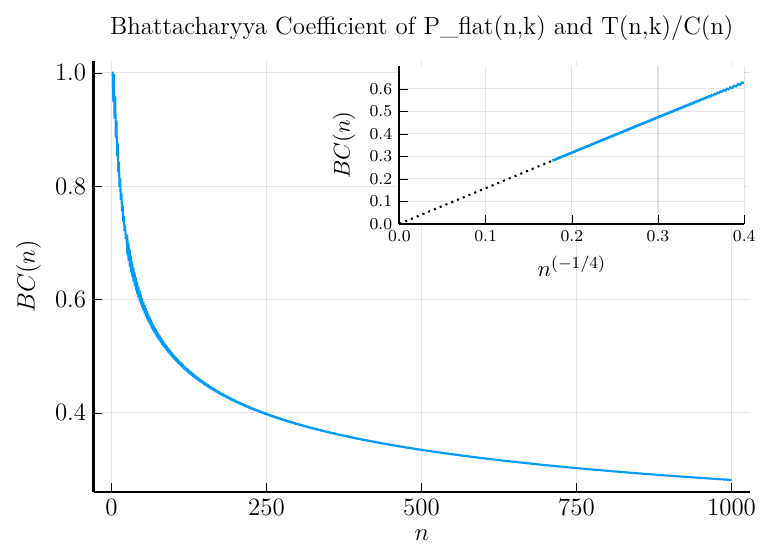}
    \caption{Plot of the Battacharyya Coefficient for the flatBias sampling and true distributions of binary trees of size $n$ relative to the number of leaves $k$. The value decays rapidly to zero implying that there is no effective overlap between the distributions for large $n$. The inset plot shows $BC(n)$ against $n^{-1/4}$ clearly demonstrating that decay has a power-law rate. The dotted black line is a manual extrapolation to guide the eye and is not intended to convey hard data.}
    \label{fig:BC_flatbias}
\end{figure}

Numerical methods confirm that this does lead to flat sampling in $k$. We once again calculate the Bhattacharyya Coefficient between the flat distribution and the true distribution with the results shown in figure~\ref{fig:BC_flatbias}. What is remarkable is that while both the original sampler and the flatBias method decay to zero as $n\to\infty$, the original method does so at an exponential rate, while the flatBias method does so only at a power law rate.

This is clearly an improvement on the unbiased sampler, but the $f(n,k) = 1$ case for all odd $n$ suggests some potential trouble. Recall that a bias function of $1$ means that for any odd $n$ we will \emph{always} choose not to increase the number of leaves. It is reasonable to question whether this will prevent us from ever reaching certain states. Indeed, while we can show that it is possible to sample all $(n,k)$ pairs, the process is no longer able to reach all configurations of binary trees. Because of this, the weight averages obtained from this process will not be estimators of the count of binary trees, but rather only the count of trees reachable using this process.

The proof that we do not reach all states is as follows. Consider some configuration of even size, $2m$, grown using the flat bias process. It must have been grown from a predecessor of size $2m-1$ by the addition of a vertex that did not increase the number of leaves. That is, it must have had a new leaf appended to the end of a previous leaf. To show that the process is not able to reach all states, it will thus suffice to provide an example of a binary tree which could not have grown in this way. An equivalent statement of this property is a configuration which has no leaves whose removal would make their parent a leaf node.

\begin{figure}[ht]
    \centering
    \includegraphics[width=0.6\textwidth]{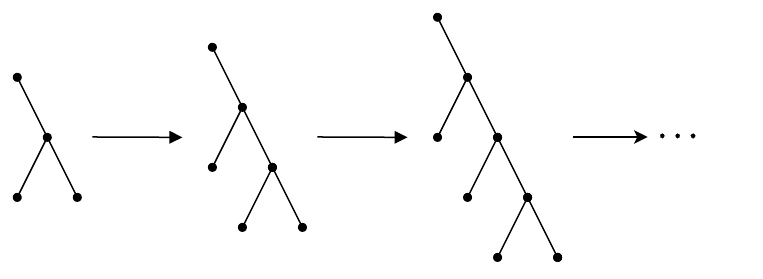}
    \caption{Examples of configurations which can not be reached using the flattening bias.}
    \label{fig:unreachable_configs}
\end{figure}

We can construct a sequence of states by taking some even-sized starting configuration and appending a pair of leaves to the right-most, lowest depth leaf. We can continue this pattern to obtain configurations of any arbitrary even size. If the first state in the sequence is chosen such that it is meets the criteria for not being sampled, having no leaves whose removal would make their parent a leaf node, then all states in this sequence will also be part of the set of unsampleable configurations. This is exactly the case of the sequence shown in figure~\ref{fig:unreachable_configs}.

We can expand on this argument to obtain a formula for the number of states which cannot be sampled by the flatBias method by means of a combinatorial decomposition. Every unsampleable state either is the unique full binary tree with three nodes, can be constructed by appending any unsampleable state to a single node, or can be build up recursively by appending any unsampleable state to either or both of the leaves of that tree. This is represented by the following diagram, where \includegraphics[height=\baselineskip]{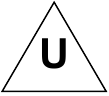} represents the set of unsampleable states:

\begin{center}
    \includegraphics[width=0.9\linewidth]{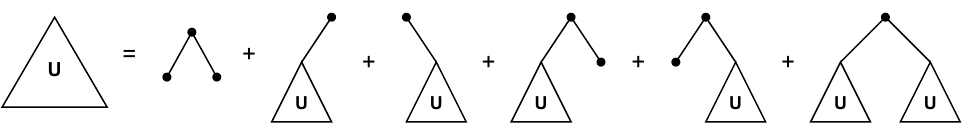}
\end{center}

Let $U(x)$ be the generating function for the sequence of the number of states which cannot be sampled by the flatBias method. The symbolic representation above can be expressed as a functional equation for $U(x)$,
\begin{equation}
    U(x) = x^3 + 2 x U(x) + 2 x^2 U(x) + x U(x)^2\;.
\end{equation}
This quadratic equation in $U(x)$ can be solved, taking the branch which reproduces the first terms of the counting sequence, to get the explicit form of the generating function
\begin{equation}
    U(x) = -\frac{2x^2 + 2x - 1 + \sqrt{8x^3-4x+1}}{2x}\;.
\end{equation}

Expanding this as a power series in $x$ give the coefficients $1,2,6,16,45,\ldots$, as expected from direct counting. This sequence can be found in the OEIS as sequence \href{https://oeis.org/A025266}{A025266}.
(with a shift due to different starting coefficients) hinting at connections to several other combinatorial objects.

The dominant singularity of the generating function is at $\sqrt{5}/4 - 1/4 \approx 0.30902$, so we can say that the asymptotic growth rate of the number of configurations of size $n$ missed by the flatBias method is approximately $3.2361^n$ to leading order for $n$ large. As discussed previously, the total number of configurations of size n is given by the Catalan numbers, which are widely known to have asymptotic growth rate $4^n$, so the proportion of missed configurations decays exponentially fast in the asymptotic limit. 

From this we can conclude that while we do miss exponentially many configurations, we can expect our weights to converge to the Catalan numbers with some exponentially decaying systematic error which will be dominated by Monte Carlo error as we sample larger and larger configurations. Despite this, if we were to attempt to sample in some non-constant energy landscape, then we would not be able to say necessarily that the missed samples do not make a significant contribution to the overall weight averages. This is compounded by the fact that the unsampled configurations are not just some random subset of binary trees, but rather a specific class of objects with common properties. In the case where the chosen weighted ensemble heavily favours this class of configuration we would not expect to obtain any useful results. 

In summary, the flat sampling method by biasing offers a unique approach to achieving uniform sampling in leaf count. While numerical methods affirm its efficacy in flat sampling with respect to $k$, the method does exhibit limitations, particularly in its inability to sample all configurations of binary trees. 

\section{Atmospheric Flattening: A New Approach}\label{sec:flattening}
A key conclusion from our diagnostics of GARM (with or without pruning and enrichment) is that the sampler is simply unable to sufficiently explore the space of possible atmospheres for the models studied. Because of this, the number of samples required for the mean weight converge to the true value is unreasonably large. 

This is very similar to problems faced when sampling some interacting models of linear polymers, where there are very hard-to-reach configurations which nonetheless make significant enough contributions that their absence in sampling will cause a misrepresentation of the ensemble average. A technique called FlatPERM has shown remarkable success in allowing one to sample these difficult models~\cite{flatPERM}, but even here there are limitations. For example, when simulating models with bimodal distributions, such as in~\cite{krawczyk_prellberg_bad_sampling}, it is very difficult to reliably sample both peaks.

In FlatPERM one chooses some additional parameters and collects the weights into bins based on the values of those parameters for the particular configuration. When making pruning and enrichment decisions, the number of copies is calculated from the current average weight for the specific bin.

Perhaps the most important consideration in designing an effective FlatPERM implementation is choosing the parameters that most thoroughly capture the complexity of the configuration space. For GARM the two obvious parameters to choose regardless of the model are the positive and negative atmospheres, as these are the primary contributors to the dynamics of the algorithm. We modify the pruning and enrichment step to calculate the enrichment ratio, $r$ (previously defined in Eqn.~(\ref{eq:enrichment_ratio}), as
\begin{equation}
r = \frac{W({\{\varphi_0\dots\varphi_n\}})}{ \hat{W}_{n, a_+, a_-}}
\end{equation}
where
\begin{equation}
\hat{W}_{n, a_+, a_-}
\end{equation}
is now the average sampled weight of configurations with size $n$, restricted to atmospheres of size $a_+$ and $a_-$. We refer to this technique as atmospheric flattening of GARM.

\subsection{Results and Limitations of Atmospheric Flattening}

We once again sampled the four models of branched polymers, this time using GARM with atmospheric flattening. For the tree-like animals, we sampled configurations up to a target size of $N=200$, but due to limitations of the method that will be discussed later, we were only able to obtain data for the non-tree-like animals up to $N=50$, and we were limited to $10,000$ tours for all four models.

\begin{figure}[ht]
    \centering
    \begin{subfigure}{0.45\textwidth}
        \includegraphics[width=\linewidth]{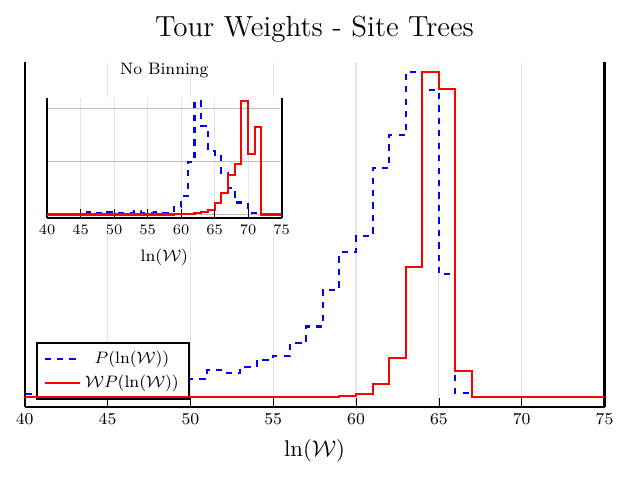}
    \end{subfigure}
    \begin{subfigure}{0.45\textwidth}
        \includegraphics[width=\linewidth]{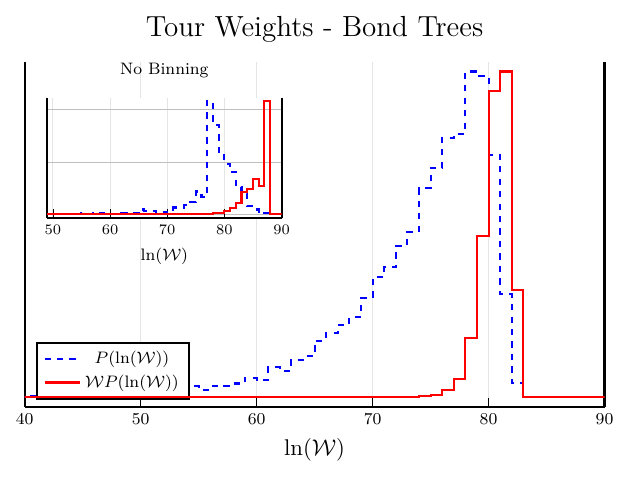}
    \end{subfigure}
    \begin{subfigure}{0.45\textwidth}
        \includegraphics[width=\linewidth]{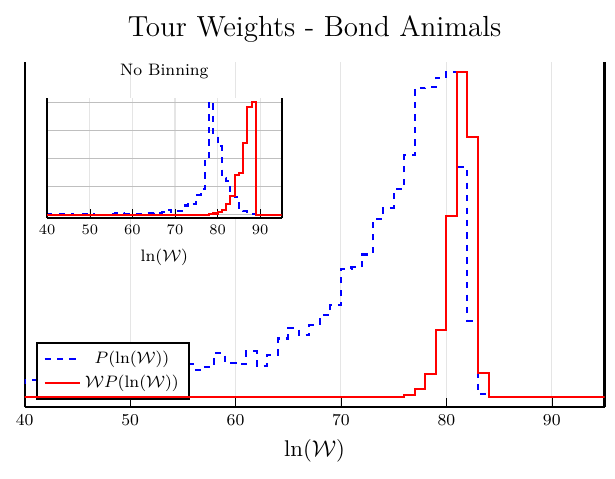}
    \end{subfigure}
    \begin{subfigure}{0.45\textwidth}
        \includegraphics[width=\linewidth]{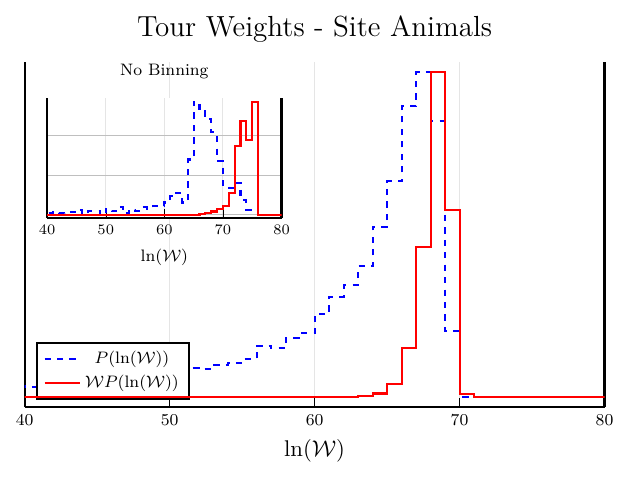}
    \end{subfigure}
    
    \caption{Diagnostic tools inspired by~\cite{hsu_grassberger_review}. The dashed blue line is a histogram of $P(\ln(\mathcal{W}))$. The solid red line is the weighted distribution, $\mathcal{W}P(\ln(\mathcal{W}))$. It is quite clear that the atmospheric flattening has improved the distribution of tour weights and so we would expect to have more reliable results. All plots show data for $N=50$.}
    \label{fig:binned_diagnostics}
\end{figure}

In figure~\ref{fig:binned_diagnostics} we present the same diagnostic plots as in section~\ref{sec:garm_results} using data obtained from runs of GARM with atmospheric flattening up to size $N=50$. The inset plots show the distributions for Pruned/Enriched GARM replicated from figure~\ref{fig:toursize_diagnostics} for comparison.

These diagnostics show that atmospheric flattening gives a clear improvement in the overlap of the tour weight distributions, suggesting that individual tours are far less likely to dominate the average.

\begin{figure}[ht]
    \centering    
    \begin{subfigure}{0.24\textwidth}
        \includegraphics[width=\linewidth]{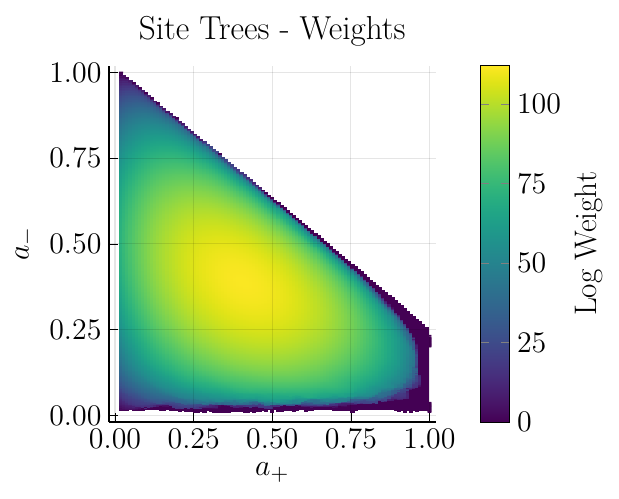}
    \end{subfigure}
    \begin{subfigure}{0.24\textwidth}
        \includegraphics[width=\linewidth]{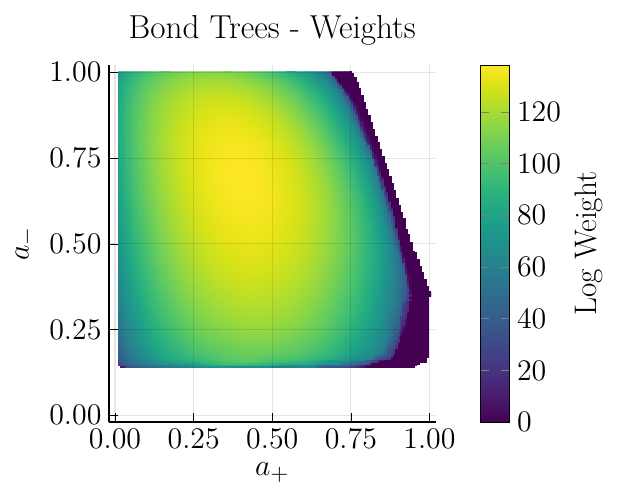}
    \end{subfigure}
    \begin{subfigure}{0.24\textwidth}
        \includegraphics[width=\linewidth]{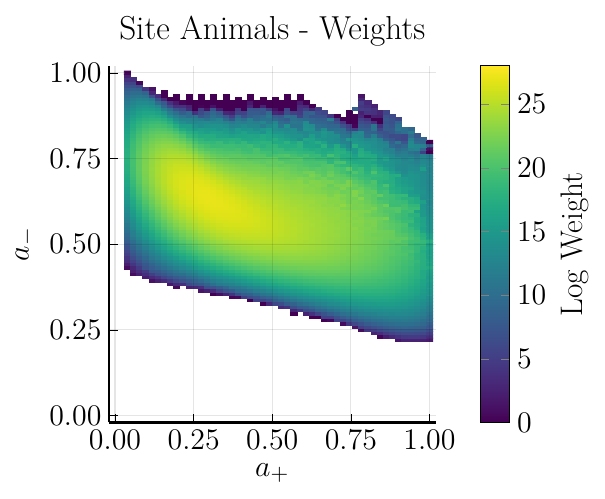}
    \end{subfigure}
    \begin{subfigure}{0.24\textwidth}
        \includegraphics[width=\linewidth]{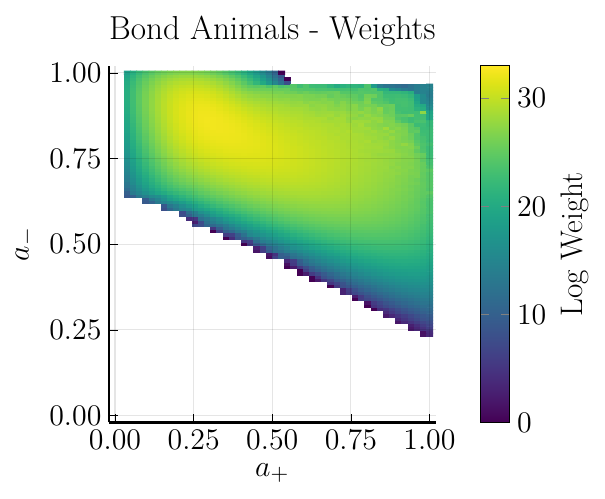}
    \end{subfigure}
    \begin{subfigure}{0.24\textwidth}
        \includegraphics[width=\linewidth]{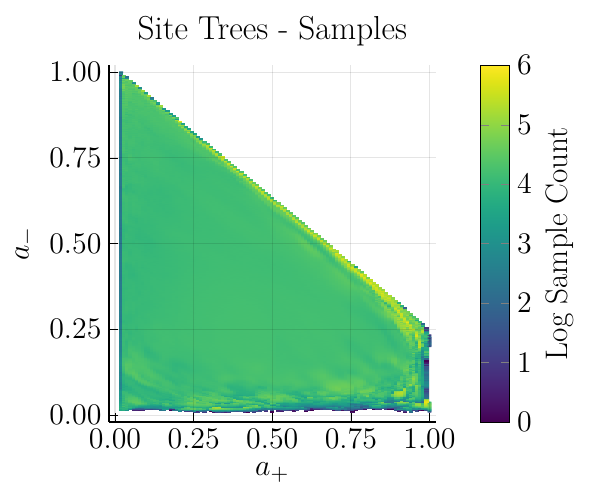}
    \end{subfigure}
    \begin{subfigure}{0.24\textwidth}
        \includegraphics[width=\linewidth]{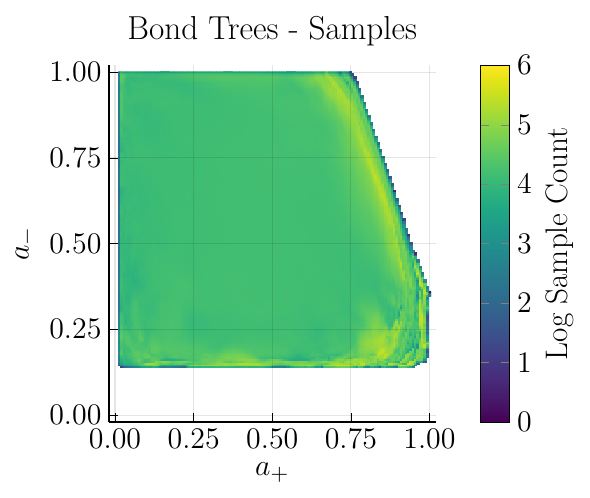}
    \end{subfigure}
    \begin{subfigure}{0.24\textwidth}
        \includegraphics[width=\linewidth]{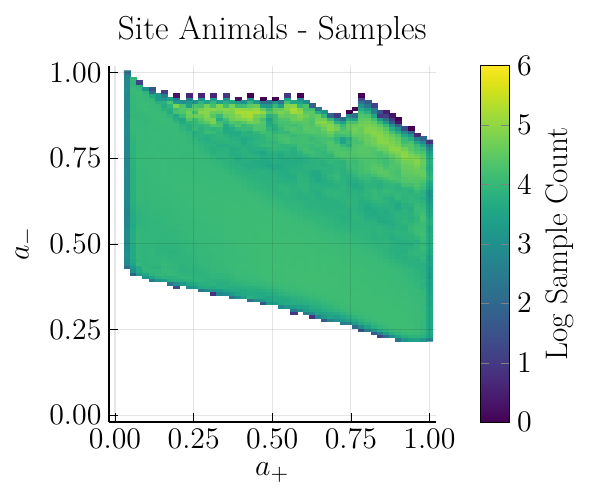}
    \end{subfigure}
    \begin{subfigure}{0.24\textwidth}
        \includegraphics[width=\linewidth]{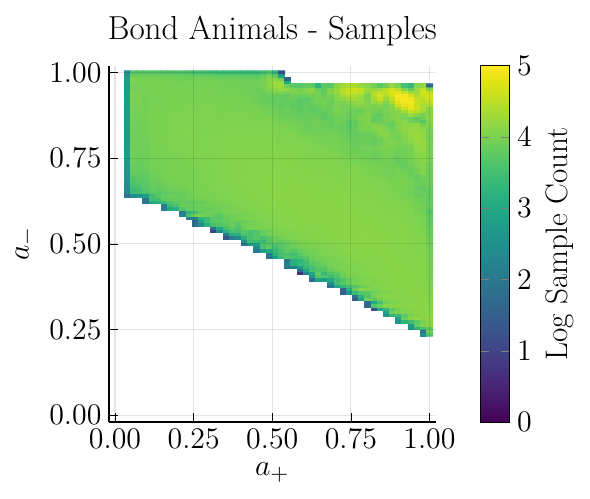}
    \end{subfigure}
    \caption{Here we show the contours for total weight and sample counts for each of the four models using the atmospheric flattening technique. For each of the four models the algorithm has almost fully explored the state space, with a near-constant number of samples across the whole distribution. The colour scale differs for each plot and is chosen to highlight the structure of the distribution, such as the location of the peaks. The data is for $N=200$ for the tree models and $N=50$ for the cyclic models.}
    \label{fig:binned_contours_allmodels}
\end{figure}

\begin{figure}[ht]
    \centering
    \begin{subfigure}{0.4\textwidth}
        \includegraphics[width=\linewidth]{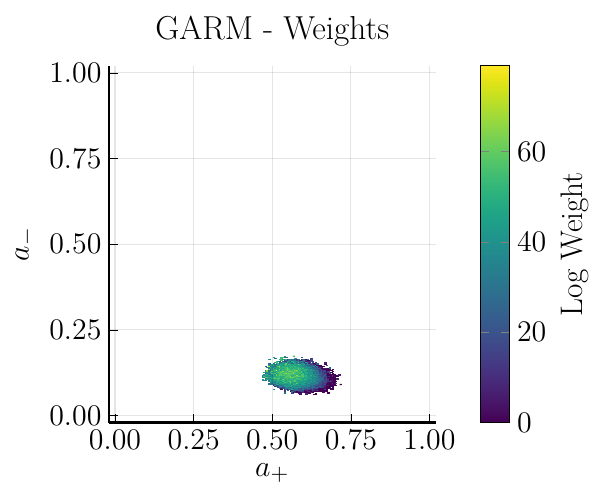}
    \end{subfigure}
    \begin{subfigure}{0.4\textwidth}
        \includegraphics[width=\linewidth]{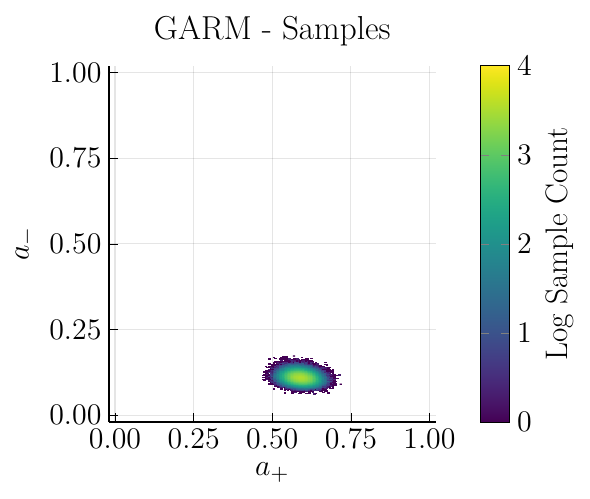}
    \end{subfigure}
    \begin{subfigure}{0.4\textwidth}
        \includegraphics[width=\linewidth]{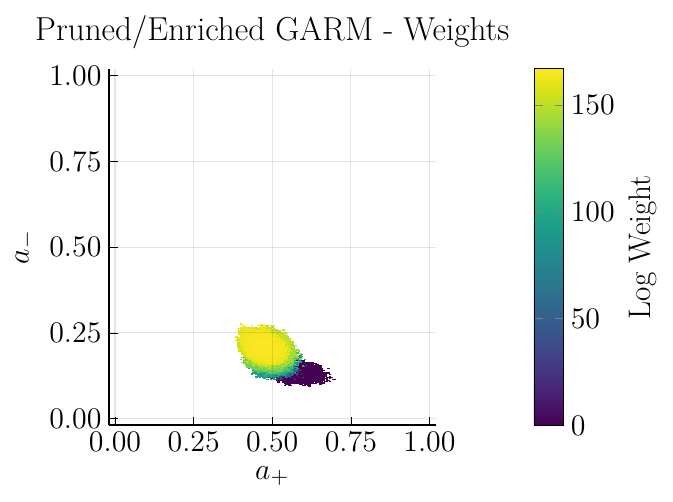}
    \end{subfigure}
    \begin{subfigure}{0.4\textwidth}
        \includegraphics[width=\linewidth]{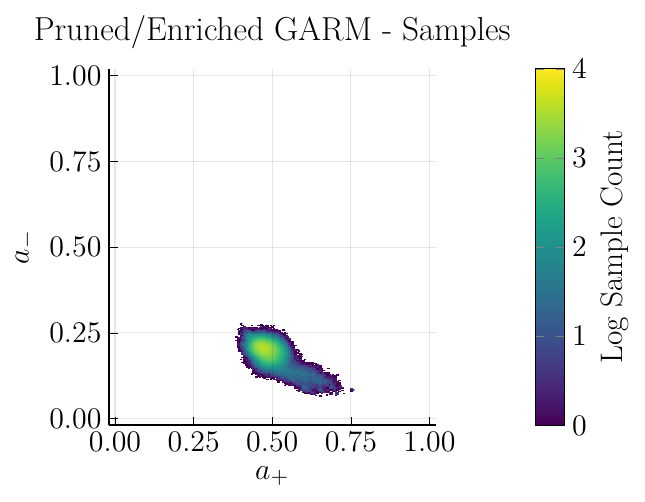}
    \end{subfigure}
    \begin{subfigure}{0.4\textwidth}
        \includegraphics[width=\linewidth]{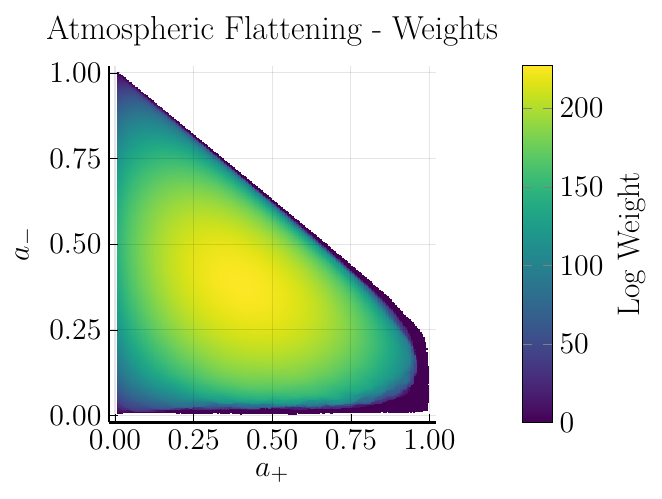}
    \end{subfigure}
    \begin{subfigure}{0.4\textwidth}
        \includegraphics[width=\linewidth]{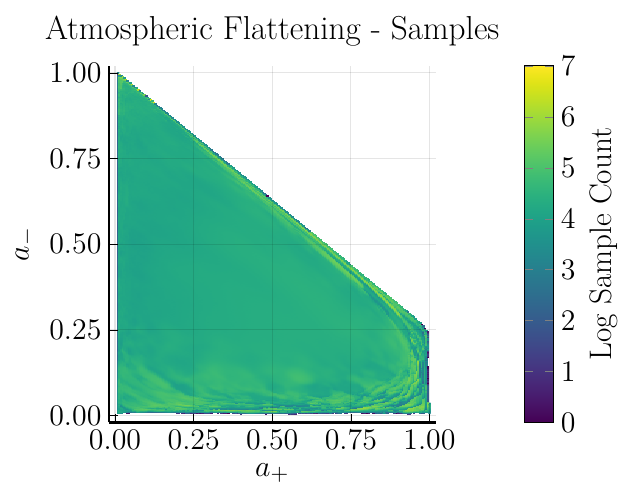}
    \end{subfigure}
    \caption{Contours for the site tree model (size $N=400$) showing the difference between plain GARM, GARM with pruning and enrichment, and GARM with atmospheric flattening in that order from top to bottom. The left column is the distribution of weights and the right column is the distribution of samples. Now that the full distribution is clear from the flattened contour, we can see just how poorly placed the samples from the first two strategies overlap with the bulk of the true weights. It is no wonder why these runs produced such poor results.}
    \label{fig:binned_contours_sitetree}
\end{figure}

In figure ~\ref{fig:binned_contours_allmodels}, we show contour plots for all four models,
obtained using atmospheric flattening. It is clear in these contours that we are much more thoroughly exploring a wide range of atmospheres. The distribution of samples is very nearly uniform, and the peak of the weights distribution sits well centered within this region. This would suggest that we are adequately sampling from the bulk of the distribution, and, unlike for the prior two methods, can expect reliable estimates of the counting numbers.

In figure~\ref{fig:binned_contours_sitetree} we compare the weight and sample distributions for all three methods, restricted to site trees. The distributions for plain GARM and Pruned/Enriched GARM are the same as those in figure~\ref{fig:pegarm_contours}), and have been reproduced without cropping to help with comparison. Looking at the figure, the problem with the original sampling distributions becomes quite apparent. The absence of pruning and enrichment resulted no samples of configurations with weights within even several orders of magnitude of the highest weight configurations. This highlights the critical role of pruning and enrichment in guiding the sampling process towards more statistically relevant configurations. Despite the improvements from pruning and enrichment, the peak of the sample distribution still falls well outside the main concentration of weights. Although not presented here due to space constraints, as we delve into larger objects, the aforementioned effect becomes more pronounced, leading to a compelling explanation for the systematic decline observed in our growth constant estimates beyond $N = 50$ (figure~\ref{fig:mu_estimates}). 

These diagnostics provide a compelling argument that atmospheric flattening should improve the ability of GARM to sample the four lattice animal models, and to validate this we return to growth constant estimates. If this new method is effective, then we will see the regression window estimates converge as the window grows, ideally to a value near the current best estimates.

\begin{figure}[ht]
    \centering
    \begin{subfigure}{0.45\linewidth}
        \includegraphics[width=\linewidth]{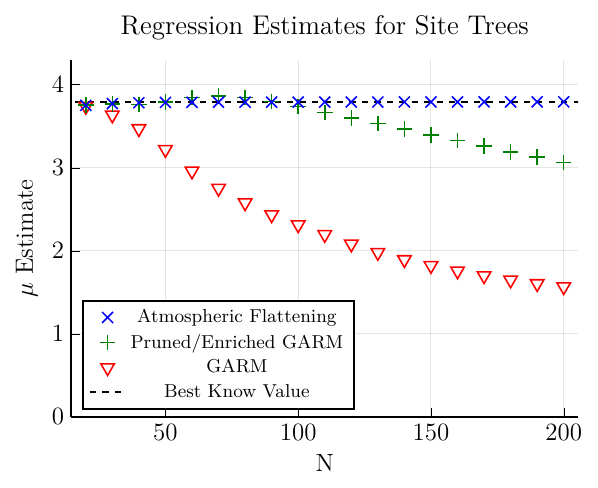}
    \end{subfigure}
    \hfill
    \begin{subfigure}{0.45\linewidth}
        \includegraphics[width=\linewidth]{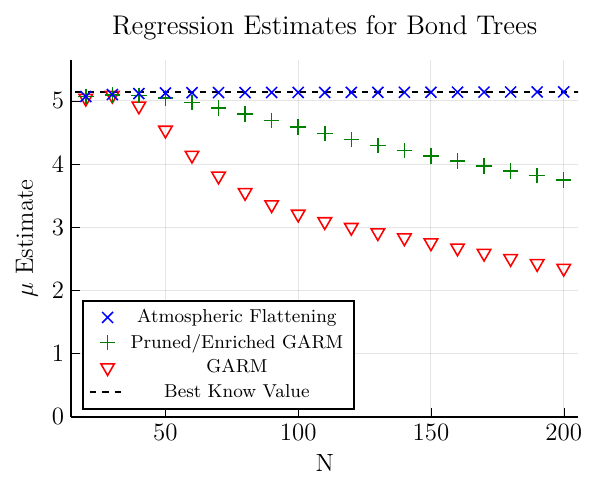}
    \end{subfigure}
    \caption{Growth constant estimates from runs using atmospheric flattening. The best known value is shown as a dotted line. These estimates are far better than those from figure \ref{fig:mu_estimates}, supporting the claim that atmospheric flattening can allow GARM to effectively sample from these models. Unfortunately due to the worsened time complexity we are not able to obtain results for non-tree models for $N>50$. We choose not to include those plots as they are largely uninformative.}
    \label{fig:mu_estimates_binned}
\end{figure}

In figure~\ref{fig:mu_estimates_binned} we replicate the data from figure~\ref{fig:mu_estimates} with the addition of a new series produced using the same regression technique on data from GARM with atmospheric flattening. The new results show excellent agreement to the current best known values, and show no apparent signs of tailing-off even up to $N=200$.

With this evidence we believe it is safe to conclude that atmospheric flattening is an effective tool for sampling lattice animal models using stochastic growth algorithms. That is not to say, however, that it does not present its own challenges. First among them is a worsened scaling of algorithm run-time with target size. We can break down the time complexity of the algorithm in terms of the number of tours run, the number of samples taken in a tour,  and the complexity of taking a single sample. 

At a given size, $n$, there is a bin for each possible $(a_+, a_-)$ pair. For all of the models studied the range of atmospheres is $O(n)$, so the total number of bins is $O(n^2)$. As we take samples at each $n$ from $1$ to our target size, $N$, the number of bins filled in each tour will be $O(N^3)$ on average. With atmospheric flattening, the number of samples taken is roughly constant in each bin, so the number of samples in each tour will be $O(N^3)$ on average. While still polynomial, this scaling severely hampered our ability to obtain data for even moderate target sizes. This was especially true for the two models with cycles, as we could obtain at best an average case $O(n)$ time per sample due to the need to calculate the articulation points at each step leading to an overall $O(N^4)$ average case time complexity per tour.

Given that a run of $100,000$ tours of site animals up to size $50$ took several days, we would expect sampling up to size $100$ to take multiple months of wall-time. This is also not a problem that can easily be alleviated by throwing more compute cores at the problem, as the depth-first nature of each tour significantly limits the speedup possible from parallelisation. Since the goal of this investigation was to provide estimates for the asymptotic growth constants, being practically limited to such small sizes of objects reduces the competitiveness of our algorithm.

The issue of computational complexity is not exclusive to atmospheric flattening but is prevalent in any flatPERM implementation involving a high number of binning parameters, such as in the simulation of pulling adsorbing and collapsing polymers from a surface \cite{krawczyk_pulling_polymers}, where the four-dimensional parameter space restricted the lengths that could be practically reached to the order of $N=100$, compared to the $N=1,000,000$ that Grassberger was able to obtain for unconstrained SAWs. 

\section{Conclusion}\label{sec:conclusions}

In this study, we undertook a comprehensive examination of the GARM sampling technique for branched polymer models. It became evident that while GARM has been invaluable in certain applications, its efficiency diminishes when applied to branched polymers. Our investigations highlighted that GARM consistently underestimates growth constants, especially as object size increases. Such discrepancies are not only alarming but point to a need for a more refined or alternative approach when studying branched polymers.

We turned our attention to simpler, rooted binary trees to dissect the challenges further. These trees, though structurally less complex, exhibited similar sampling issues, solidifying our concerns regarding the limitations of GARM. However, this exploration also provided an avenue for improvement. By biasing the transition probabilities we were able to direct GARM to sample in the correct locations.

Drawing parallels from issues faced in sampling certain interacting models of linear polymers, we considered the FlatPERM technique, which has been successful in efficiently these more complex models. By adapting FlatPERM to the GARM approach and focusing on parameters like the positive and negative atmospheres, we aimed to capture the complexities inherent in the configuration space.

Our modifications, while promising, highlighted the intricate balance between model complexity and sampling efficiency. Our empirical findings underscore the need for a more nuanced understanding of the dynamics involved in GARM, especially in the realm of branched polymers.

\section*{Acknowledgements}
We thank Andrew Rechnitzer for helpful comments, in particular the suggestion of studying GARM on binary trees. This research utilised Queen Mary's Apocrita HPC facility, supported by QMUL Research-IT. \href{http://doi.org/10.5281/zenodo.438045}{http://doi.org/10.5281/zenodo.438045}

\section*{References}

\bibliographystyle{unsrt}


\end{document}